\documentclass[lettersize,journal]{IEEEtran}
\usepackage{amsmath,amsfonts}
\usepackage{adjustbox}
\usepackage{array}
\usepackage{algorithmic}
\usepackage{multirow}
\usepackage[caption=false,font=normalsize,labelfont=sf,textfont=sf]{subfig}
\usepackage{textcomp}
\usepackage{stfloats}
\usepackage{caption}
\usepackage{xcolor} 
\usepackage{array}       
\usepackage{makecell}    
\usepackage{soul}
\usepackage{pifont}     
\usepackage{xcolor}     
\usepackage{array}
\usepackage{makecell}
\usepackage{cite}
\usepackage[ruled,vlined]{algorithm2e}
\IEEEoverridecommandlockouts

\usepackage{url}
\usepackage{verbatim}
\usepackage{graphicx}
\begin{document}

\title{O-RAN xApps Conflict Prediction using Graph Convolutional Networks}

\author{Maryam Al Shami,~\IEEEmembership{Graduate Student Member and WIE Member,~IEEE}, Jun Yan,~\IEEEmembership{Member,~IEEE}, and Emmanuel Thepie Fapi
}
\markboth{}%
{Shell \MakeLowercase{\textit{et al.}}: A Sample Article Using IEEEtran.cls for IEEE Journals}


\maketitle

\begin{abstract}
Open Radio Access Network (O-RAN) adopts a flexible, open, and virtualized architecture with standardized interfaces, reducing dependency on a single supplier. O-RAN hosts many intelligent applications known as eXtended Applications (xApps). xApps are applications deployed at the RAN Intelligent Controller (RIC) that leverage advanced Artificial Intelligence/Machine Learning (AI/ML) algorithms to make dynamic decisions for network optimization. Each application operates with distinct optimization objectives and is managed by independent operators while accessing shared network resources. Conflicts in this context occur when a deployed xApp's objective interferes with another xApp, resulting in incompatible actions or decisions that may negatively impact network performance.
The lack of a unified mechanism to coordinate and prioritize the actions of different applications can create three types of conflicts (direct, indirect, and implicit). Conflict prediction in O-RAN refers to the proactive analytical process through which potential interactions or behaviors that may lead to conflicts between network applications are identified in advance, prior to their manifestation within the operational system.
In our paper, we introduce a novel data-driven Graph Convolutional Network (GCN)-based method called GRAPH-based Intelligent xApp Conflict Prediction and Analysis (GRAPHICA). It predicts three types of conflicts (direct, indirect, and implicit) and pinpoints the root causes (xApps). GRAPHICA captures the complex and hidden dependencies among the xApps, controlled parameters, and key performance indicators (KPIs) in O-RAN to predict possible conflicts. Then, it identifies the root causes (xApps) contributing to the predicted conflicts. 
The proposed method is evaluated using highly imbalanced synthetic datasets, in which conflict instances constitute between 40\% and merely 10\% of the data. This evaluation setting is designed to reflect realistic operational environments where conflicts are infrequent, thereby enabling a comprehensive assessment of the model’s performance under real-world conditions.
Experimental results demonstrate a high F1-score over 98\% for the synthesized datasets with different levels of class imbalance.
\end{abstract}


\section{Introduction}
\IEEEPARstart{T}{he} Open Radio Access Network (O-RAN) represents a paradigm shift in how mobile networks are developed and operated. Traditionally, mobile networks are designed 
such that each layer of the cellular protocol stack is implemented as a black box with a limited number of vendors. This limits RAN reconfigurability and coordination among nodes, making it difficult to deploy and integrate equipment from multiple vendors \cite{Understanding_O-RAN_2023}. Motivated by this, O-RAN is introduced to decouple the network's hardware and software components. Thus, promoting interoperability, openness, and flexibility in the design and deployment of RAN components \cite{Understanding_O-RAN_2023}. 

Service Management and Orchestration (SMO) and Radio Intelligent Controller (RIC) are two important components of the O-RAN architecture. 
The SMO is responsible for the overall orchestration, management, and automation of the O-RAN \cite{Understanding_O-RAN_2023}. 
The RIC serves as a platform to deploy applications that use machine learning models to predict network behavior, detect anomalies, and enhance decision-making capabilities \cite{RIC_2021}. The intelligent, data-driven control with the RICs is provided through two logical controllers: Non-Real Time (Non-RT) RIC and Non-Real Time (Near-RT) RIC \cite{Intelligence_and_Learning_in_O-RAN_for_Data-Driven_NextG_Cellular_Networks_2021}. These two RICs provide a platform to host third-party applications that could be powered by machine learning to orchestrate the RAN for specific tasks \cite{O-RAN_A_Concise_Overview_2025}. 
They differ in terms of the functions they perform and the timescales over which they operate. 

In this context, eXtended Applications (xApps) are software applications designed to run on the Near-RT RIC. These applications are designed to make real-time, fast-paced decisions that have an immediate impact on day-to-day operations and user experiences. This allows a flexible and dynamic control over various aspects of the network behavior, including load balancing \cite{xApp_for_Traffic_Steering_and_Load_Balancing_2023}, interference management \cite{Interference_Mitigation_with_ML_Based_xAPP_2024}, and handover optimization \cite{Handover_O-RAN_2024,Mobility_Management_O-RAN_2024}. 

The autonomous operation of xApps managed by different third-party vendors makes them susceptible to conflicts. Each xApp has its own objective and decision-making process. This will introduce new challenges as these applications independently interact with the network environment. The actions of one or more applications could potentially adversely impact the objectives of the others as they share the same network resources or parameters \cite{Challenges_for_conflict_mitigation_2023,RIC_Apps_Conflict_Management_i14y_lab_2024,Open_RAN_xApps_Design_and_Evaluation_2024}. 

Artificial Intelligence and Machine Learning (AI/ML) algorithms have been leveraged to address xApp conflicts management in O-RAN \cite{Team_Learning_Based_Resource_2022,A_Game_Theoretic_Approach_2023,QACM_QoS_Aware_xApp_Conflict_Mitigation_2024,Learning_and_Reconstructing_Conflicts_in_O-RAN_2024}. Nevertheless, these approaches frequently depend on distributional assumptions, most notably the presumption that the underlying data follows a Gaussian distribution for deriving KPI values. In practice, however, KPI behavior is often considerably more complex and may deviate substantially from such assumptions \cite{QACM_QoS_Aware_xApp_Conflict_Mitigation_2024}.  

Additionally, existing methods in the literature often overlook the inherently imbalanced nature of real-world datasets. In practical O-RAN deployments, data distributions across different network conditions, user behaviors, or anomaly types are rarely uniform, leading to significant challenges in model training and performance generalization \cite{Predictive_alarm_models_for_improving_radio_access_network_2025}. Most current approaches do not explicitly address class imbalance or the prediction of rare events. This oversight limits the applicability of such methods in operational environments, where the ability to handle data imbalance is crucial for achieving reliable and fair decision-making.

Motivated by the need for a more comprehensive understanding of interdependencies within the O-RAN, GCN has demonstrated strong performance across various domains to model the complex nonlinear relationships between elements. This includes power load prediction, drug synthesis, few-shot learning, and link
prediction \cite{Fault_diagnosis_of_power_transformers_2021}. GCN offers key advantages such as scalability and generalization, enabling efficient modeling of complex, structured data and adapting to varying network topologies \cite{GCN_model_2016, Survey_of_Graph_Neural_Network_for_Internet_of_Things_and_NextG_Networks_2025}. Additionally, GCN excels at learning structural representations over graphs and is well-suited for aggregating and reasoning over time-dependent, multi-entity interactions than DRL-based approaches \cite{Combining_Deep_Reinforcement_Learning_With_Graph_Neural_Networks_2021}. 

By modeling these elements as nodes within a graph and their interactions as edges, GCNs are capable of learning intricate dependencies across the network. This representation allows the model to understand how actions taken by individual xApps influence system-wide behavior, enabling the real-time prediction of potential conflicts. Moreover, the graph-based insights learned by the GCN can be used to pinpoint the specific xApps responsible for performance issues, thereby facilitating accurate RCA and supporting proactive conflict resolution. This becomes especially valuable when analyzing high-dimensional time-series data, as is common in commercial RANs, where thousands of KPIs are collected every few minutes.


Accordingly, we propose GRAPHICA, a GCN-based method that is capable of handling highly imbalanced datasets with up to 10\% conflicts. It predicts the xApps conflicts (direct, indirect, and implicit) and identifies their root causes (xApps) for future mitigation strategies. The core of our approach lies in leveraging a binary-state dataset that captures the dynamic behavior of xApps, controllable parameters, and KPIs, based on their changes relative to the previous time instance. The decision to use binary state representations instead of relying on Gaussian distribution assumptions is driven by practical considerations observed in real-world KPI datasets. Many KPIs exhibit stochastic behavior, which can limit the applicability of distribution-based approaches \cite{Adaptive_Thresholding_Heuristic_for_KPI_2024}. In operational O-RAN systems, this results in large volumes of time-series data across the network, with potentially subtle and nonlinear dependencies that emerge only through temporal interaction patterns. In such settings, rule-based or heuristic methods quickly become intractable, brittle, or incapable of generalization, particularly predicting indirect or implicit conflicts, where causality is not local and may span multiple hops across the graph.


The main contributions of this paper are summarized as
follows:
\begin{itemize}
    \item We propose a 5G O-RAN xApps conflict prediction with a root cause analysis method using Graph Convolutional Networks (GCNs). This allows us to capture the complex and hidden dependencies between the xApps, parameters, and KPIs in the network to predict xApps conflicts (direct, indirect, and implicit).
    \item We leverage a binary state-based GCN structure to map and explain the complex relationships. This structure enables conflicts prediction while offering key inputs for the root cause analysis module, which can be extended to other cellular networks. 
    
    
    \item We use the focal loss function in the model training. This allows the model to handle highly imbalanced datasets. The model achieves an F1 score greater than 98\% for the synthesized datasets with different levels of class imbalance.
\end{itemize}

The remainder of this paper is structured as follows. 
Section \ref{sec:problem_formulation} presents the problem formulation, clearly defining the research question and outlining the scope and objectives of the study. Section \ref{sec:proposed_method} describes the proposed system model. Section \ref{sec:experiments_and_results} details the evaluation metrics, the experimental setup, and the interpretation of the results. Section ~\ref{sec:qualitative_analysis} discusses the results derived from the analysis. Section \ref{sec:conclusion} concludes this paper.

\section{Related Work}
Conflict management in O-RAN is of critical importance, as the coexistence of multiple xApps with overlapping control objectives can lead to inconsistent or suboptimal network behavior. To address this challenge, existing solutions increasingly employ AI/ML techniques for conflict detection, prevention, and mitigation \cite{A_Software_Defined_Radio_based_O_RAN_Platform_for_xApp_Conflict_Detection_and_Mitigation_2024,xApp_Level_Conflict_Mitigation_in_O_RAN_a_Mobility_Driven_Energy_Saving_Case_2025,Conflict_Mitigation_Approach_for_O_RAN_xApps_2025,Conflict_Mitigation_Framework_and_Conflict_Detection_2023}. 


In this section, we categorize the related work on detecting or predicting conflict occurrences in O-RAN into two distinct groups. For each category, we critically examine the key limitations and highlight how our work differs from these approaches. Table~\ref{table: comparison_with_other_methods} shows a comparison between GRAPHICA and other methods in terms of the use case, results, distribution independence, scalability, and the imbalance nature of the dataset under study \cite{Team_Learning_Based_Resource_2022, A_Game_Theoretic_Approach_2023, QACM_QoS_Aware_xApp_Conflict_Mitigation_2024, Learning_and_Reconstructing_Conflicts_in_O-RAN_2024, PACIFISTA_2024, xapp_distillation_2024}. 


\begin{table*}[h]
\centering
\caption{Comparison with Other Methods}
\begin{adjustbox}{width=\textwidth}
\begin{tabular}{|>{\centering\arraybackslash}m{2.6cm}|>{\centering\arraybackslash}m{1cm}|>{\centering\arraybackslash}m{3cm}|>{\centering\arraybackslash}m{5cm}|>{\centering\arraybackslash}m{1.6cm}|>{\centering\arraybackslash}m{1.6cm}|>{\centering\arraybackslash}m{1.6cm}|}
\hline
\textbf{Method} & \textbf{Dataset} & \textbf{Use Case} & \textbf{Results} & \textbf{Distribution Independent} & \textbf{Scalable} & \textbf{Imbalanced Dataset} \\ \hline        
Team learning \cite{Team_Learning_Based_Resource_2022} & Synthetic & Power allocation and radio resource allocation & 

\begin{itemize}
    \item 8\% higher throughput and 64.8\% lower packet drop rate
\end{itemize}
 
& \ding{55} & \ding{55} & \ding{55}  \\ \hline
Game theory \cite{A_Game_Theoretic_Approach_2023} & Synthetic & General & 

\begin{itemize}
    \item Mitigation of conflicts between xApps 
\end{itemize}

& \ding{55} & \ding{55} & \ding{55}  \\ \hline
QoS-Aware Conflict Mitigation \cite{QACM_QoS_Aware_xApp_Conflict_Mitigation_2024} & Synthetic & General & 

\begin{itemize}
    \item Effective in maintaining QoS requirements for conflicting xApps
\end{itemize}

& \ding{55} & \ding{51} & \ding{55}  \\ \hline
xApp Distillation \cite{xapp_distillation_2024} & Synthetic & General & 

\begin{itemize}
    \item Consistent 10 Mbps downlink data rate 
\end{itemize}

& \ding{51} & \ding{55} & \ding{55}  \\ \hline
PACIFISTA \cite{PACIFISTA_2024} & Synthetic & Throughput maximization and energy saving &  

\begin{itemize}
    \item Effective in characterizing conflicts and providing insights for informed
xApps deployment decision  
\end{itemize}

& \ding{51}  & \ding{55} & \ding{55}  \\ \hline
GraphSAGE \cite{Learning_and_Reconstructing_Conflicts_in_O-RAN_2024} & Synthetic  & General &  

\begin{itemize}
    \item 88\% reconstruction accuracy, 100\% detection of conflicts  
\end{itemize} 

& \ding{55} & \ding{55} & \ding{55}  \\ \hline
\textbf{GRAPHICA} & Synthetic & General & 

\begin{itemize}
    \item F1-score: \{40\% conflicts: 0.9930, 30\% conflicts: 0.9946, 20\% conflicts: 0.9908, 10\% conflicts: 0.9954\}  
\end{itemize} 

& \ding{51} & \ding{51} & \ding{51}  \\ \hline

\end{tabular}
\label{table: comparison_with_other_methods}
\end{adjustbox}
\end{table*}

\subsection{Non-graph-based methods}
\begin{itemize}
    \item Deep Q-learning methods: 
    In \cite{Team_Learning_Based_Resource_2022}, a team deep Q-learning algorithm for resource allocation is disclosed to mitigate xApps conflicts in O-RAN. However, it faces scalability challenges when applied to a larger number of xApps
    \cite{Conflict_Mitigation_Framework_and_Conflict_Detection_2023}.
    This could lead to a "state space explosion" 
    to explore all possible states and actions effectively \cite{A_Deep-Q_Learning_Scheme_2022}. 
    Besides, it
    does not explicitly focus on capturing structural relationships between entities or nodes \cite{A_Comprehensive_Discussion_on_Deep_Reinforcement_Learning_2021}.
    In \cite{xapp_distillation_2024}, a Deep Q-Network and a multi-headed multilayer perceptron network model are presented.
    Relying on a single distilled model 
    necessitates repeated distillation as network conditions change over time \cite{xApp_Conflict_Mitigation_with_Scheduler_2025}. In this paper, we propose the use of GCNs due to their scalability and flexibility, which enable efficient modeling of complex dependencies among data and adaptability to dynamic network topologies. Moreover, our approach eliminates the need for explicit state exploration.


    \item Game-theoretic methods: In \cite{A_Game_Theoretic_Approach_2023}, a bargaining game-theoretic approach is introduced. 
    Integrating game-theoretic models into decentralized systems makes it difficult to make optimal decisions with incomplete information, especially when leader nodes are unable to observe the strategies of follower nodes \cite{Game_theoretic_Designs_for_Blockchain-based_IoT_2022}.
    This work is further improved in \cite{QACM_QoS_Aware_xApp_Conflict_Mitigation_2024}, proposing a cooperative game theory. 
    Nonetheless, the validation of this method is limited to Python-based simulations using a simplified network model, which may not fully capture the complexities of real-world KPMs prediction \cite{PACIFISTA_2024}. Unlike game-theoretic approaches, our proposed method offers a significant advantage in decentralized systems. GCN enables the effective aggregation of local information across nodes, facilitating optimal decision-making even in the presence of incomplete information. Additionally, it leverages structural dependencies within the network to infer hidden information, resulting in enhanced adaptability in dynamic environments.
    

    \item Other methods: Another study \cite{Detection_and_mitigation_of_indirect_conflicts_between_xApps_2023} presents a technique for identifying indirect conflicts between xApps in O-RAN. 
    Although this work provides valuable insights, it addresses indirect conflict solely and ignores direct and implicit conflicts.
    The authors in \cite{COMIX_2025} propose two DRL-based xApps to detect conflicts between xApps in O-RAN using a network digital twin (NDT).
    However, the need for an NDT to evaluate both proposed actions and every xApp-generated value of network control parameters could lead to 
    questioning the practicality of depending on pretrained xApps \cite{xApp_Conflict_Mitigation_with_Scheduler_2025}. 
    In contrast, we propose a framework that predicts the three xApps conflict types (direct, indirect, and implicit). It is an event-based approach that allows our model to generalize effectively across diverse KPI contexts without being constrained to specific distributions.

    \end{itemize}
    
\subsection{Graph-based methods} Other works address the xApps conflict management using graph-based approaches. 

\begin{itemize}
    \item Hierarchical graphs: The authors in \cite{PACIFISTA_2024} propose a method to detect conflicts among applications in O-RAN. Accordingly, they combine hierarchical graphs with statistical knowledge of applications to find dependencies between control parameters and Key Performance Measurements (KPMs). The main challenge involves developing detailed statistical profiles for each application across various operational scenarios and in the context of interactions with multiple other applications \cite{xApp_Conflict_Mitigation_with_Scheduler_2025}. In our work, GCN offers a distinct advantage by efficiently capturing the complex interdependencies between applications, thereby alleviating the challenge of developing detailed statistical profiles for each application across diverse operational scenarios and interactions with multiple other applications.
    \item GraphSAGE: In \cite{Learning_and_Reconstructing_Conflicts_in_O-RAN_2024}, the paper suggests using GraphSAGE for reconstructing and labeling conflict graphs in O-RAN. It captures the hidden dependencies between xApps, parameters, and Gaussian distributed KPIs. 
    Then, graph labeling is performed to identify the different types of conflicts based on well-defined, graph-based definitions. 
    The problem is formulated as a link prediction for reconstructing and labeling conflict graphs in O-RAN based on predefined graph-based conflicts. The method constructs a multivariate time-series graph representing temporal relationships, trains the model using Mean Squared Error (MSE) loss, and applies a fixed threshold to binarize correlations, thereby generating a reconstructed adjacency matrix that represents the conflict graph.
    One limitation of this study is the lack of consideration for data imbalance, 
    while assuming that KPI values follow a Gaussian distribution. Contrarily, GRAPHICA formulates the problem as a graph prediction using GCN and is capable of predicting conflicts in highly imbalanced datasets with up to 10\% conflicts. Our proposed method encodes multivariate time-series data into a binary-state dataset, capturing the behavioral changes of xApps, controllable parameters, and KPI values based on their observed dynamic patterns. From this binary representation, a graph-structured dataset is constructed, which is subsequently used to train a GCN model employing a focal loss function. Our model is distribution-independent, as KPIs are assigned binary state values based on their dynamic behavior. This design allows our approach to generalize effectively across various KPI contexts, without being limited to specific distributions.
    \end{itemize}

Although the previously discussed non-graph-based approaches offer valuable insights, 
they do not encode the relational structure among xApps, parameters, and KPIs, such as which xApp controls which parameters, or how parameters influence specific KPIs. This makes them less suitable for capturing complex interdependencies inherent in systems like O-RAN.
Without awareness of these interdependencies, the xApps operating independently without knowledge of each other's actions may unintentionally counteract each other and continuously override each other. This leads to instability or oscillation in the network behavior and degrading network performance \cite{PACIFISTA_2024}. 
Effective RCA depends not only on identifying the outcome, but also on understanding the underlying cause. This involves tracing the problem back through the decisions made by xApps, the parameters they altered, and the subsequent impact on relevant KPIs. The reviewed graph-based methods, despite their effectiveness, still fall short of fully accounting for the imbalanced nature of datasets in practical life scenarios, as these conflicts are rare. Addressing this limitation is crucial for enhancing the generalizability of the proposed approach in real-world applications.

Differing from existing work, this paper formulates the xApps conflict RCA in O-RAN as a graph-level classification problem. GRAPHICA uses a GCN-based approach to predict xApps conflicts on highly imbalanced datasets with conflict instances occurring at frequencies as low as 10\%. Our model is trained using a focal loss function to minimize the difference between the detected and true labels. Additionally, it identifies the root causes (xApps) contributing to the detected conflicts, enabling future mitigation strategies. To overcome the shortcomings observed in the existing studies, our approach is designed to be distribution-independent, assigning binary state values to KPIs based on their observed dynamic patterns. This allows our method to generalize effectively across a wide range of KPI contexts, without being restricted to any specific statistical distribution. Although the papers reviewed in the literature do not explicitly discuss the balancing of their datasets, the significant impact of imbalanced data on model performance and the need for methodologies tailored to address such challenges are well documented \cite{Predictive_alarm_models_for_improving_radio_access_network_2025}. Considering this, our model is specifically developed to handle highly imbalanced datasets, an area that has not been adequately addressed in the existing body of work.

\section{Preliminaries}
\label{sec:problem_formulation}
\begin{figure}[ht!]
\centering
    \begin{adjustbox}{width=0.5\textwidth}
        \includegraphics[width=0.5\textwidth]{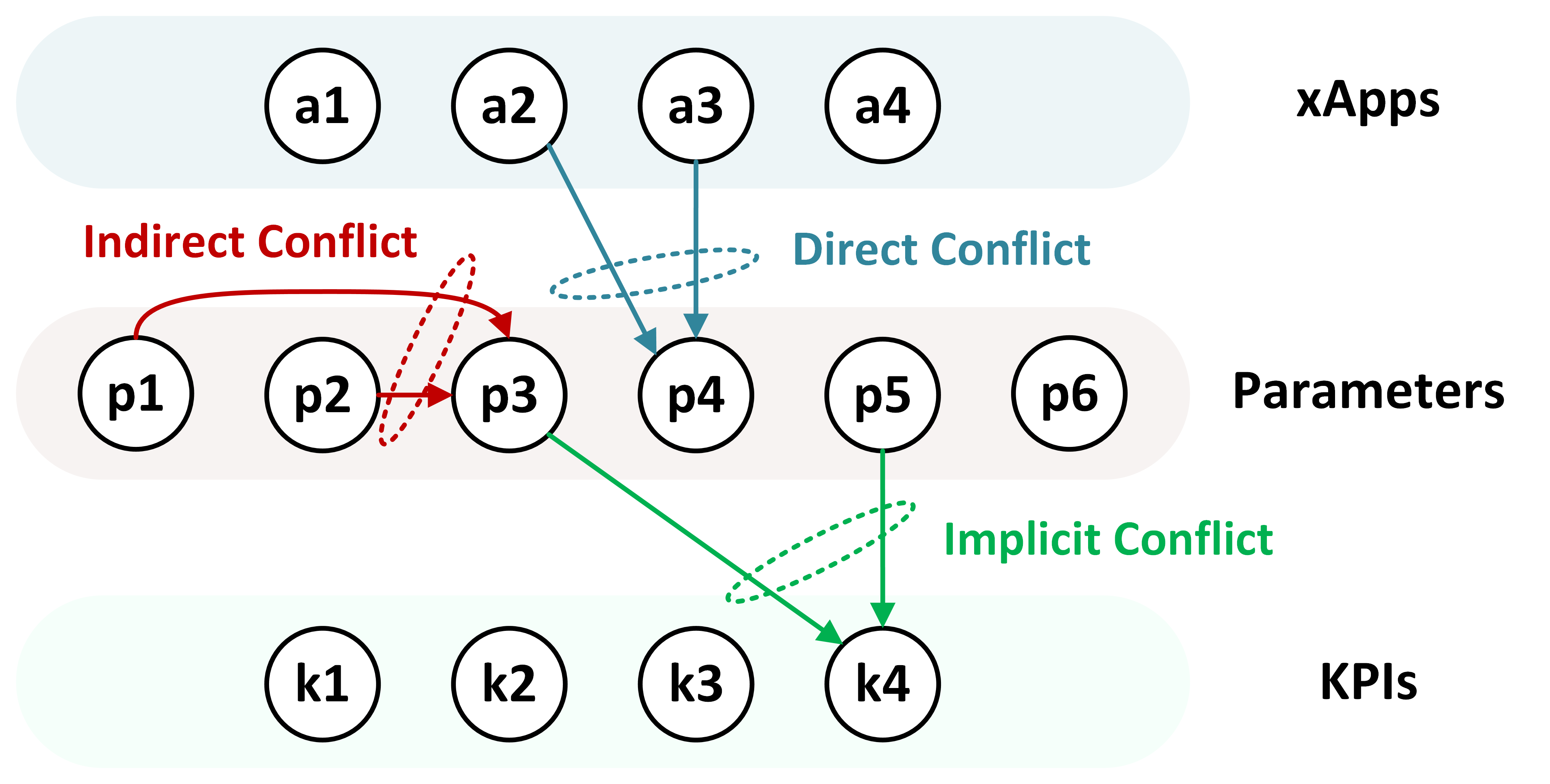}
    \end{adjustbox}
\caption{Examples of xApps conflicts.}
\label{fig: conflicts_examples}
\end{figure}

Conflict management is a crucial aspect of O-RAN, ensuring different applications operate harmoniously within the network. The conflicts can be categorized into three types: Direct, indirect, and implicit conflict \cite{Pre-Emptive_Conflict_Detection_2024,PACIFISTA_2024,xApp_Conflict_Detection_and_Mitigation_2024}. Fig.~\ref{fig: conflicts_examples} illustrates examples of the different types of conflicts that may occur within the network. In this figure, $a1$ to $a4$ represent xApps deployed in the near-RT RIC, $p1$ to $p6$ denote controllable network parameters, and $k1$ to $k4$ correspond to the observed KPIs.

Direct conflicts arise when multiple applications attempt to modify or request different settings for the same network parameter simultaneously. These conflicts are easily noticeable and can result in immediate service degradation, unstable performance, or even network failures if not promptly addressed. For example, two xApps might attempt to adjust the transmission power of the same cell simultaneously. In particular, one xApp seeks to increase power for better coverage, while the other aims to reduce it to minimize interference with neighboring cells.

Indirect conflicts are not immediately visible, but the interdependencies between the parameters and resources involved can be observed. These conflicts occur when multiple applications modify a set of parameters that directly affect the values of other parameters. They are subtler than direct conflicts and may result in unintended negative effects. For instance, one xApp may optimize load balancing by redistributing traffic across various cells, while another xApp is adjusting handover parameters to enhance mobility performance. Although these actions are independent, the resulting traffic shifts could disrupt the handover process or increase congestion in certain cells, leading to a degraded user experience.

Implicit conflicts arise when multiple xApps make control decisions that target different optimization objectives, thereby interfering with one another. These conflicts are not directly observable. They depend on intrinsic relationships between control parameters and observable KPIs. They are more abstract and arise from differences in how xApps interpret the network's requirements. For instance, one xApp might prioritize energy efficiency by reducing transmission power across several cells to lower consumption. Meanwhile, another xApp seeks to increase power in certain areas to enhance coverage. The conflicting goals, energy efficiency versus coverage, could lead to poor service quality in some areas of the network or result in inefficient power usage.

Grasping the distinctions between these conflicts in O-RAN is crucial for developing resilient, automated networks. The network must be capable of efficiently handling various optimization goals such as QoS, coverage, and energy efficiency. Effective conflict management is necessary as a tailored resolution approach to maintain optimal network performance while achieving specific objectives. 

Dealing with xApps in an O-RAN environment brings many critical challenges that need to be addressed. This involves ensuring that multiple xApps can operate simultaneously and complement each other, leading to a well-optimized network. Specifically, there is a lack of coordination and standardization among the various components responsible for RAN control decisions. Moreover, O-RAN networks often evolve due to upgrades and adjustments to meet new requirements. These continuous changes can result in frequent conflicts, making manual conflict management not only unfeasible but also unsustainable over time.

In addition, as multiple conflicting objectives can coexist, finding a single optimal solution that optimizes all objectives may not be feasible. Thus, it becomes challenging to identify the conflicts along with their root causes to address them \cite{Open_RAN_xApps_Design_and_Evaluation_2024,Challenges_for_conflict_mitigation_2023}.

Graph Neural Networks (GNNs) present a promising approach for addressing xApp conflicts in O-RAN and identifying the root causes, given their ability to effectively model complex dependencies and interactions within dynamic network environments. GNN is a graph-based method
designed to work with graph-structured data. Graphs are an ideal data structure for representing various types of real-world systems and the relationships among them
\cite{GNN_model_2009}. By leveraging the capabilities of GNNs, the problem can be traced by analyzing the decisions made by xApps, the parameters they modified, and the resulting effects on the associated KPIs.

Graph Convolutional Networks (GCNs) are a specific type of GNN that uses convolutional operations to propagate information between nodes in a graph. The node's new representation is a weighted sum of its neighbors' features, and the weight depends on the graph structure itself \cite{GCN_model_2016}. GCNs have gained widespread popularity and are extensively used in a variety of applications to capture complex dependencies within dynamic networks, such as sparse mobile crowdSensing \cite{Spatiotemporal_Urban_Inference_2023}, traffic prediction \cite{STEP_2021, DMSTG_2024}, and user identity linkage \cite{EgoMUIL_2024}. 

The large volume of multivariate time-series data in O-RAN is difficult to handle efficiently by heuristics or rule-based methods. GCNs are suitable as they can learn complex relational patterns, capture nonlinear dependencies, and scale large datasets \cite{Fault_diagnosis_of_power_transformers_2021,GCN_model_2016,Combining_Deep_Reinforcement_Learning_With_Graph_Neural_Networks_2021}. The limitations of heuristic and deep reinforcement learning approaches, which rely heavily on accurate mathematical models, are addressed by learning complex relational dependencies directly from data without requiring explicit system equations \cite{Survey_of_Graph_Neural_Network_for_Internet_of_Things_and_NextG_Networks_2025}. With multiple propagation layers, information can diffuse across the entire graph, enabling the GCN to capture higher-order dependencies and long-range interactions.

Multiple xApps may simultaneously contend for shared network resources, such as radio spectrum, computational capacity, and memory. GCNs are well-suited to model these resource dependency graphs and to predict potential conflicts arising from concurrent control actions. For instance, xApp\_Beamforming aims to optimize antenna radiation patterns to maximize throughput, whereas xApp\_PowerControl seeks to reduce transmission power to enhance energy efficiency. A conflict arises because beamforming actions typically increase power requirements, while power control simultaneously attempts to reduce transmission power, leading to incompatible control objectives. Accordingly, GCN can be a good candidate to capture those higher-order dependencies and relationships.

\section{Proposed Method}
\label{sec:proposed_method}
\begin{figure}[t!]
\centering
    \begin{adjustbox}{width=0.47\textwidth}
        \includegraphics[width=0.4\textwidth]{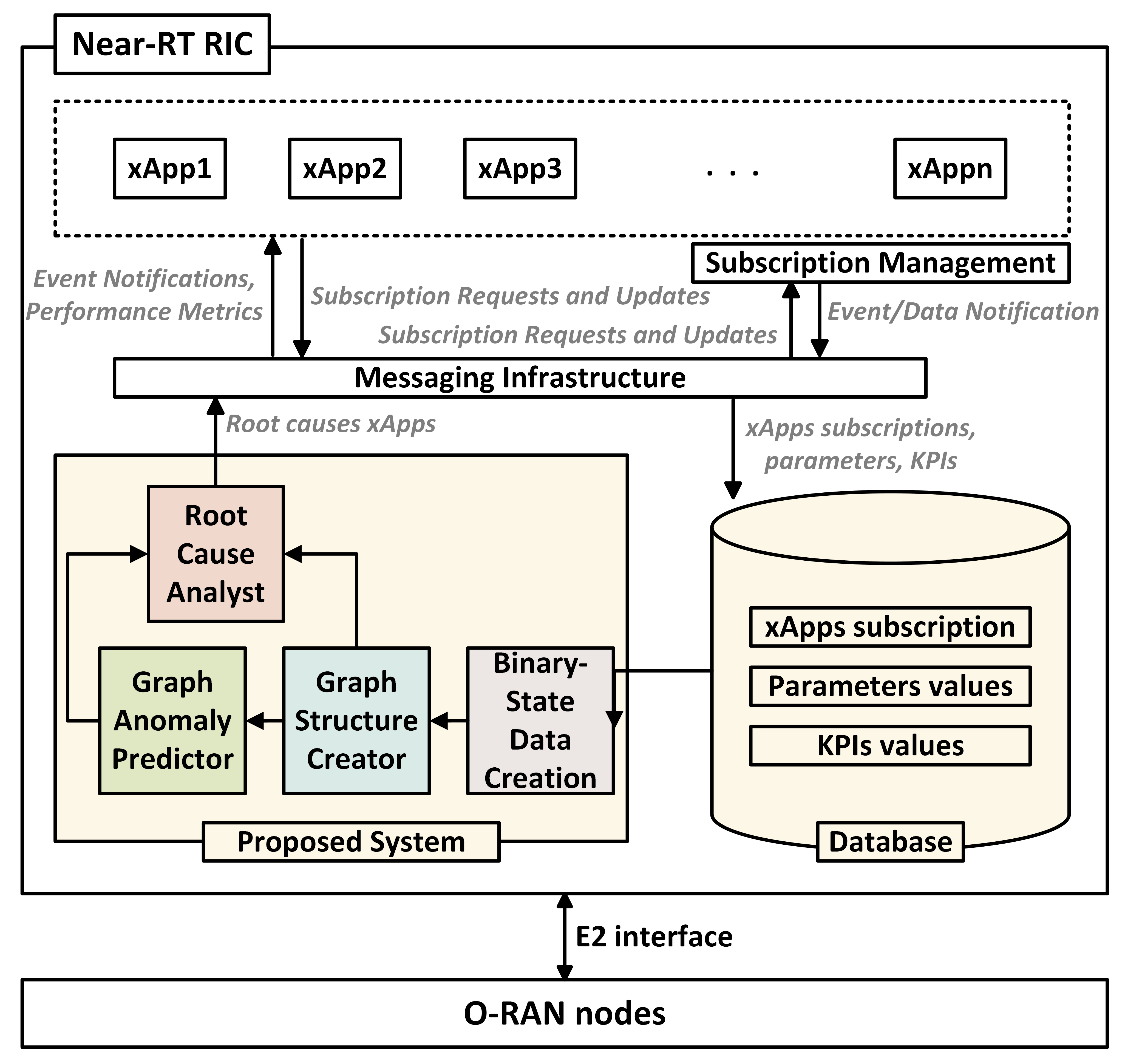}
    \end{adjustbox}
\caption{The proposed framework.}
\label{fig: proposed_framework}
\end{figure}

In the following, we introduce our proposed conflict model. The proposed framework is depicted in Fig.~\ref{fig: proposed_framework}. The proposed solution comprises three major modules: Graph Structure Creator (GSC), Graph Anomaly Predictor (GAP), and Root Cause Analyst (RCA).

\subsection{Initial Data Processing}

The Subscription Manager (SM) within the Near-RT RIC is responsible for managing and coordinating xApp subscription requests and updates. The subscription values define which xApps are associated with specific controllable parameters. The state changes of xApps, parameters, and KPIs can be retrieved through the application programming interfaces (APIs) offered by the RIC platform and can be systematically recorded in a database for subsequent analysis.

Let the collected data at time $t$ consist of $N$ data points, indexed by $i = 1, 2, \dots, N$, each encompassing xApp activation states, controllable parameters, and KPIs. The data can be formally represented as:
\[
\mathcal{D}_t^{(i)} = \{ \mathcal{A}_t^{(i)}, \mathcal{P}_t^{(i)}, \mathcal{K}_t^{(i)} \},
\]
where $\mathcal{A}_t^{(i)}$, $\mathcal{P}_t^{(i)}$, and $\mathcal{K}_t^{(i)}$ denote the xApp activation status, controllable parameters, and KPIs for data point $i$ at time $t$, respectively.

where the sets are defined as follows:
\begin{itemize}
    \item Let $\mathcal{A} = \{ a_1, a_2, a_3, \dots, a_n \}$ be the set of O-RAN xApps activation state deployed on the Near-RT RIC,  
    \item Let $\mathcal{P} = \{ p_1, p_2, p_3, \dots, p_m \}$ be the set of controllable parameters, and  
    \item Let $\mathcal{K} = \{ k_1, k_2, k_3, \dots, k_l \}$ be the set of observable KPIs in the network.
\end{itemize}

Our proposed solution necessitates the implementation of a dedicated component that transforms the collected data $\mathcal{D}$ into a binary-state dataset $\mathcal{B}$. Specifically, changes in the behavior of the data relative to the previous timestamp are encoded as binary values, as depicted in Fig.~\ref{fig: Binary Dataset Generation and GSC}.  

To construct the binary-state representation of each data point, a binary-state transformation function $f$ is required that maps the observed system state at time $t$ and data point $i$ to a structured binary vector. Specifically, for each data point $i$, the binary state is denoted as:

\begin{equation}
\mathcal{S}_t^{(i)} =
\Big\{
s_{\mathcal{A}}^{(i)}, \; s_{\mathcal{P}}^{(i)}, \; s_{\mathcal{K}}^{(i)}
\Big\},
\end{equation}

where each component is a vector of binary states corresponding to the individual elements in the set:

\begin{align}
s_{\mathcal{A}}^{(i)} &= 
\big[ s_{a_1}^{(i)}, s_{a_2}^{(i)}, \dots, s_{a_n}^{(i)} \big], \\
s_{a_j}^{(i)} &=
\begin{cases}
1, & a_{j,t}^{(i)} \neq a_{j,t-1}^{(i)}, \\
0, & \text{otherwise},
\end{cases} 
\quad j = 1,2,\dots,n,
\end{align}

\begin{align}
s_{\mathcal{P}}^{(i)} &= 
\big[ s_{p_1}^{(i)}, s_{p_2}^{(i)}, \dots, s_{p_m}^{(i)} \big], \\
s_{p_j}^{(i)} &=
\begin{cases}
1, & p_{j,t}^{(i)} \neq p_{j,t-1}^{(i)}, \\
0, & \text{otherwise},
\end{cases} 
\quad j = 1,2,\dots,m,
\end{align}

\begin{align}
s_{\mathcal{K}}^{(i)} &= 
\big[ s_{k_1}^{(i)}, s_{k_2}^{(i)}, \dots, s_{k_l}^{(i)} \big], \\
s_{k_j}^{(i)} &=
\begin{cases}
1, & k_{j,t}^{(i)} \neq k_{j,t-1}^{(i)}, \\
0, & \text{otherwise},
\end{cases} 
\quad j = 1,2,\dots,l.
\end{align}

\begin{equation}
\mathcal{S}_t^{(i)} = \{ s_{\mathcal{A}}^{(i)}, \; s_{\mathcal{P}}^{(i)}, \; s_{\mathcal{K}}^{(i)} \}.
\end{equation}

Accordingly, the binary-state dataset for data point $i$ can be expressed as a collection of component-wise binary vectors:

\begin{equation}
\mathcal{B}_t^{(i)} = f\big(\mathcal{D}_t^{(i)}, \mathcal{D}_{t-1}^{(i)}\big) 
= \{ s_{\mathcal{A}}^{(i)}, \; s_{\mathcal{P}}^{(i)}, \; s_{\mathcal{K}}^{(i)} \}, 
\quad i = 1,2,\dots,N,
\end{equation}

where each element in $s_{\mathcal{A}}^{(i)}$, $s_{\mathcal{P}}^{(i)}$, and $s_{\mathcal{K}}^{(i)}$ takes a value in $\{0,1\}$, representing whether the corresponding xApp, parameter, or KPI has changed relative to the previous timestamp.

\subsection{Binary-State Dataset Creation}
The lack of publicly available datasets for conflict management motivates us to create a synthesized dataset that simulates real-world scenarios where conflicts are rare. The dataset synthesis process follows an event-driven approach rather than a time-based sampling methodology. Consequently, no fixed sampling rate or duration is defined for data generation, as instances are constructed based on the occurrence and characterization of specific events rather than periodic data collection. We will thoroughly explain the dataset generation and the modeling of the three types of conflict steps. 

Let \(\mathcal{A} = \{ \text{a}_1, \text{a}_2, \text{a}_3, \dots, \text{a}_{10}\}\)
, \(\mathcal{P} = \{ \text{p}_1, \text{p}_2, \text{p}_3, \dots, \text{p}_{15}\}\)
, and \(\mathcal{K} = \{ \text{k}_1, \text{k}_2, \text{k}_3, \dots, \text{k}_{20}\}\). 


Conflict modeling is conducted by randomly pairing elements from each set with those in the other set based on the conflict definition in Fig.~\ref{fig: conflicts_examples}, without any predefined sequence or structure. The primary objective of this approach is to simulate potential conflicts by exploring diverse combinations/dependencies of O-RAN xApps, parameters, and KPIs, fostering a comprehensive understanding of interaction dynamics in the system.


In particular, three functions are implemented, $``create\_dict\_A\_P"$, $``create\_dict\_K\_P"$, and $``create\_dict\_P'\_P"$, which generate three dictionaries: $D_{AP}$, $D_{KP}$, and $D_{P'P}$, respectively. These dictionaries establish relationships between elements of the three sets: $\mathcal{A}$ (xApps), $\mathcal{P}$ (controllable parameters), and $\mathcal{K}$ (KPIs). 

$\mathcal{P}$ and $\mathcal{P'}$ represent two sets with identical elements to model the interactions between the parameters and the influence exerted when two or more parameters affect the behavior of another parameter. The dictionary $``dict\_P'\_P"$ maps a parameter in $\mathcal{P}$ to one or more parameters in $\mathcal{P'}$. The mapping ensures logical relationships similar to the other functions (shared values, unique assignments, etc.) and explicitly avoids any case where a parameter is mapped to itself.

The values and keys from the input lists are selected using uniform random sampling without replacement. This method guarantees that each element has an equal probability of being chosen, without the application of weighting, thereby promoting diversity in the assignments.



\begin{figure*}[t!]
\centering
    \begin{adjustbox}{width=\textwidth}
        \includegraphics[trim=0cm 0cm 0cm 0cm, width=\textwidth]{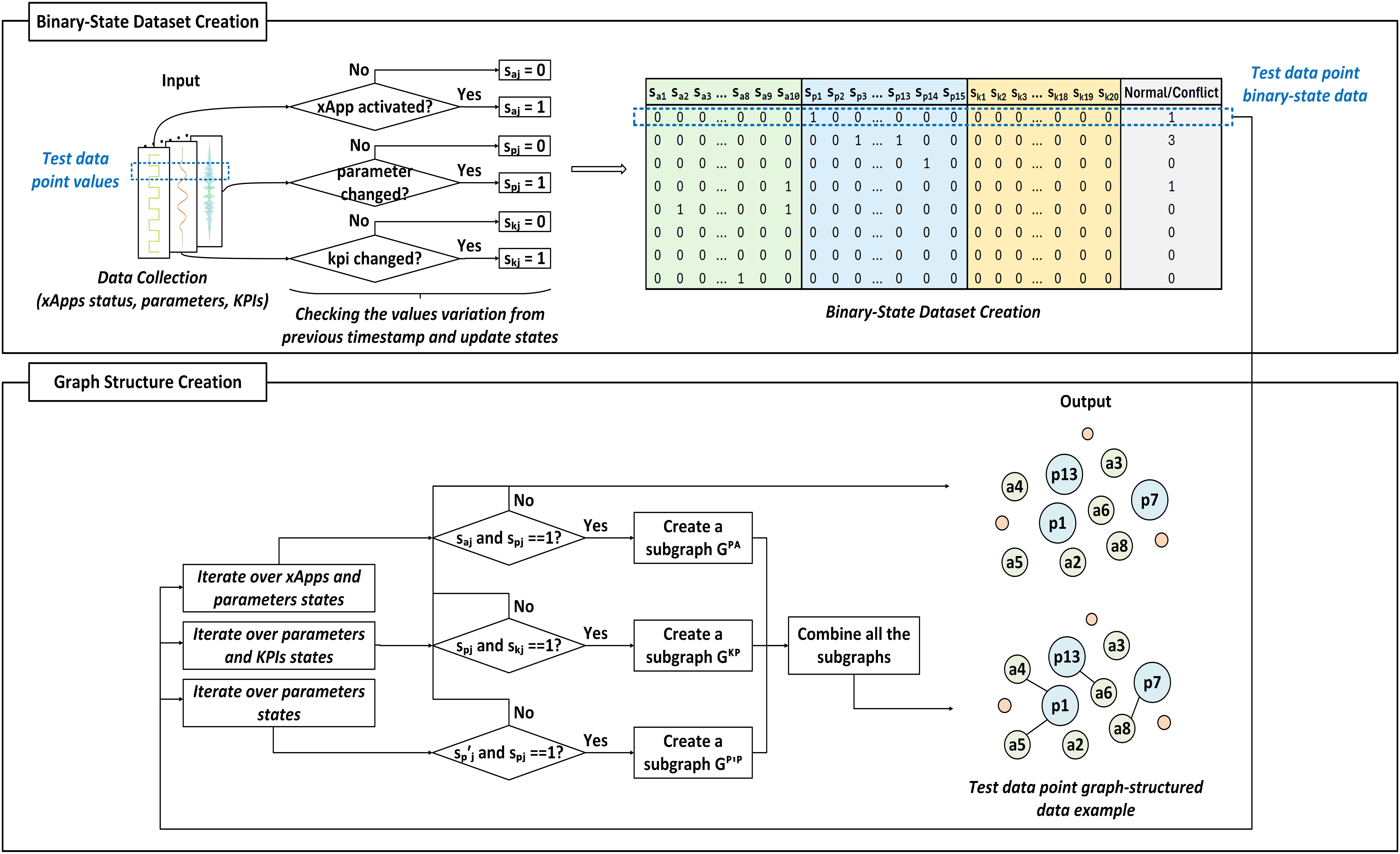}
    \end{adjustbox}
\caption{The binary-state data creation and graph structure creation modules steps.}
\label{fig: Binary Dataset Generation and GSC}
\end{figure*}

The created datasets for this study are structured such that each row corresponds to a distinct test data point, encapsulating the states of the xApps, controllable parameters, and KPIs at a timestamp t as shown in Fig.~\ref{fig: Binary Dataset Generation and GSC}. These states are represented by binary values according to the mapping used to model the conflicts, indicating the dynamic behavior of the xApp, parameter, or KPI. For instance, if two xApps are mapped to the same parameter, the state of the two xApps and the parameter will be set to ``1". The target label for each data point indicates the nature of the network's condition, whether the instance is classified as ``normal" (0: normal) or indicative of a conflict (1: direct conflict, 2: implicit conflict, and 3: indirect conflict). The detailed balanced dataset generation and conflicts modeling pseudocode is shown in Algorithm 1.



In our proposed setup, the influence of a node (whether an xApp, controllable parameter, or KPI) on another is drawn from a pool of nodes, with a constraint that no more than three nodes can influence the same target node. This design choice is consistent with other studies and reflects realistic operational scenarios, as noted in \cite{PACIFISTA_2024,QACM_QoS_Aware_xApp_Conflict_Mitigation_2024,A_Game_Theoretic_Approach_2023}. However, since our dataset is binary-based rather than value-based, as used in previous works, a greater diversity of node combinations is required to effectively capture interdependencies between nodes in a supervised manner, facilitating conflict prediction. The selected configuration, comprising 10 xApps, 15 controllable parameters, and 20 KPIs, is designed to enhance the diversity of event combinations, aligning with the event-based nature of our model. This enables the model to capture a broader range of interactions more effectively, emphasizing scenario-based occurrences over fixed-interval time sampling.

To assess the performance of the proposed model, we begin by constructing a balanced dataset comprising an equal number of instances from all four classes: 0 (normal), 1 (direct conflict), 2 (implicit conflict), and 3 (indirect conflict). Subsequently, to investigate the model's performance under class imbalance, specifically with varying proportions of normal instances (e.g., 60\%, 70\%, 80\%, and 90\%), we generate imbalanced datasets according to the following procedure:

\begin{itemize}
\item The dataset is first partitioned into four subsets, each corresponding to one of the four classes.
\item The desired percentage of samples for each class is defined based on the target class distribution.
\item A random subset of data is drawn from each class to match the specified proportions.
\item The sampled subsets are then concatenated and randomly shuffled to form the final dataset.
\end{itemize}

This process results in multiple skewed datasets with varying degrees of class imbalance, which are useful for benchmarking model applicability and evaluating performance under realistic conditions where the minority classes are underrepresented.



\begin{algorithm}
\caption{Binary-State Dataset Creation}
\begin{algorithmic}[1]
\renewcommand{\algorithmicrequire}{\textbf{Input:}}
\renewcommand{\algorithmicensure}{\textbf{Output:}}
\REQUIRE Sets $\mathcal{A} = \{a_1,\dots,a_n\}$, $\mathcal{P} = \{p_1,\dots,p_m\}$, $\mathcal{K} = \{k_1,\dots,k_l\}$; Number of events $N$
\ENSURE Dictionaries $\{ D_{AP}^{(i)}, D_{KP}^{(i)}, D_{P'P}^{(i)} \}$; Dataset $\{ \mathcal{B}_t^{(i)} \}$

\FOR{$i = 1$ to $N$}
    \STATE $D_{AP}^{(i)} \gets$ \texttt{create\_dict\_A\_P}($\mathcal{A}, \mathcal{P}$)
    \STATE $D_{KP}^{(i)} \gets$ \texttt{create\_dict\_K\_P}($\mathcal{P}, \mathcal{K}$)
    \STATE $\mathcal{P'} \gets$ copy of $\mathcal{P}$
    \STATE $D_{P'P}^{(i)} \gets$ \texttt{create\_dict\_P'\_P}($\mathcal{P'}, \mathcal{P}$), disallow $p=p'$

    \STATE Assign values randomly among the sets ($\mathcal{A}, \mathcal{P}$), ($\mathcal{P}, \mathcal{K}$), and ($\mathcal{P'}, \mathcal{K}$)
    \STATE Enforce logic constraints:
    \STATE (1) $\geq$ 2 keys share a value
    \STATE (2) $\geq$ 2 keys have unique multi-values
    \STATE (3) $\geq$ 2 keys with overlapping lists
    \STATE (4) $\geq$ 2 keys with unique values only

    \STATE Initialize $s_{\mathcal{A}}^{(i)} \in \{0,1\}^n$, $s_{\mathcal{P}}^{(i)} \in \{0,1\}^m$, $s_{\mathcal{K}}^{(i)} \in \{0,1\}^l$

    \FOR{$a_j \in \mathcal{A}$}
        \STATE $s_{a_j}^{(i)} = 1$ if $a_j$ is a key in $D_{AP}^{(i)}$
    \ENDFOR

    \FOR{$p_j \in \mathcal{P}$}
        \STATE $s_{p_j}^{(i)} = 1$ if $p_j$ is a key or value in any of $D_{AP}^{(i)}, D_{KP}^{(i)}, D_{P'P}^{(i)}$
    \ENDFOR

    \FOR{$k_j \in \mathcal{K}$}
        \STATE $s_{k_j}^{(i)} = 1$ if $k_j$ is a value in $D_{KP}^{(i)}$
    \ENDFOR

    \STATE Define $\mathcal{S}^{(i)} = \{ s_{\mathcal{A}}^{(i)}, s_{\mathcal{P}}^{(i)}, s_{\mathcal{K}}^{(i)} \}$

    \STATE Compute label $y^{(i)} \in \{0,1,2,3\}$ based on relationships and conflict type

    \STATE Append $(\mathcal{S}^{(i)}, y^{(i)})$ to dataset
\ENDFOR

\RETURN $\{ D_{AP}^{(i)}, D_{KP}^{(i)}, D_{P'P}^{(i)} \}$ and dataset $\{ \mathcal{B}_t^{(i)} \}$
\end{algorithmic}
\end{algorithm}

\subsection{Graph Structure Creator (GSC)} 
The GSC module is designed to ingest the binary-state dataset and create graph-structured data (see Fig.~\ref{fig: Binary Dataset Generation and GSC}). In our framework, intricate relationships among nodes are not known a priori. Instead, the Graph Structure Creator (GSC) module constructs a graph based on simultaneous state and behavioral changes observed in high-dimensional, multivariate time-series data. The GSC establishes connections solely according to concurrent co-variations in node behavior, without any knowledge of whether these interactions influence each other, nor any information regarding the propagation of conflict effects across the system. This module is essential, as GCN, the core model for the subsequent module in our proposed solution, is a specialized class of neural networks that operates on graph-structured data to learn hierarchical representations for graph-based predictions.


The GSC module categorizes the binary-state test data into three groups based on the number of xApps, controllable parameters, and KPIs involved. The first category includes the states of xApps and parameters, the second category focuses solely on the states of the parameters, and the third category encompasses both the states of the parameters and the KPIs. The goal of the three groups defined by the GSC module is to capture and represent different levels of interaction among network entities, xApps, controllable parameters, and KPIs, based on their binary state transitions. Specifically, the first group ($G^{\mathrm{PA}}$) aims to model the joint behavior of xApps and their associated parameters, highlighting how application-level decisions affect control settings. The second group ($G^{\mathrm{P'P}}$) isolates the dynamics of the controllable parameters alone, enabling focused analysis on parameter-level changes. The third group ($G^{\mathrm{KP}}$) integrates the states of both parameters and KPIs to reveal the impact of parameter adjustments on network performance. By constructing subgraphs for each of these categories and linking nodes based on simultaneous state changes, the GSC module enables the system to generate a comprehensive and structured graph representation of system behavior, which is then used for conflict prediction.

For each category, the GSC iterates through the states and constructs a subgraph by creating links between nodes whose behaviors (states) change simultaneously. The proposed approach is event-driven, meaning that conflict prediction occurs immediately when an xApp is activated, a parameter changes, or a KPI value is updated. This real-time response allows the framework to predict conflicts as soon as relevant events happen, effectively capturing both simultaneous and delayed conflicts through the continuous monitoring of state changes. This results in the creation of three subgraphs: $G^{\mathrm{PA}}$, representing the xApps and parameters; $G^{\mathrm{KP}}$, representing the parameters and KPIs; and $G^{\mathrm{P'P}}$, representing the controllable parameters. These three subgraphs are subsequently merged into a single graph. If the behavior of the xApps, parameters, and KPIs remains unchanged from the previous timestamp, the resulting graph will contain no edges, indicating no state transitions. The three subgraphs are merged into a single unified graph to enable the GCN model to learn from the full spectrum of interdependencies among xApps, controllable parameters, and KPIs within a single, holistic representation. While each subgraph captures a distinct aspect of the system’s behavior, xApp-to-parameter interactions, parameter-to-parameter dependencies, and parameter-to-KPI influences, analyzing them in isolation would limit the model’s ability to learn how these components jointly contribute to system state changes or conflicts. By merging the subgraphs, the GCN can propagate information across all node types and edge categories simultaneously, allowing it to capture multi-faceted, cross-layer correlations. This integration ensures that the model is not only aware of localized behavior (e.g., between parameters) but also how changes in one part of the system cascade through others, ultimately affecting performance indicators. Assigning category-specific edge attributes allows the GCN to distinguish the origin of each relationship during convolution, preserving the semantic context of each edge while enabling unified graph processing.


Let \( f_v \) denote the nodal feature and \( f_e \) denote the edge attribute for a given subgraph \( G \). The values of \( f_v \) and \( f_e \) are assigned based on the subgraph category as follows:

\begin{equation}
f_v = f_e =
\begin{cases}
1, & \text{if } G = G^{\mathrm{PA}} \quad \\
2, & \text{if } G = G^{\mathrm{KP}} \quad \\
3, & \text{if } G = G^{\mathrm{P'P}} \quad 
\end{cases}  
\end{equation}

GCN uses the node features as the starting point to update the node embeddings (representations) through multiple layers of convolutions. These embeddings are refined iteratively as the model propagates information across neighboring nodes. Including edge attributes in the graph convolution process allows the model to learn a more nuanced understanding of node interactions. When performing convolution, edge attributes help the GCN understand the nature of the relationship between nodes. The detailed GSC pseudocode is shown in Algorithm 2.


\subsection{Graph Anomaly Predictor (GAP)}
\begin{figure*}[t!]
\centering
    \begin{adjustbox}{width=\textwidth}
        \includegraphics[trim=0cm 1cm 0cm 0cm, width=\textwidth]{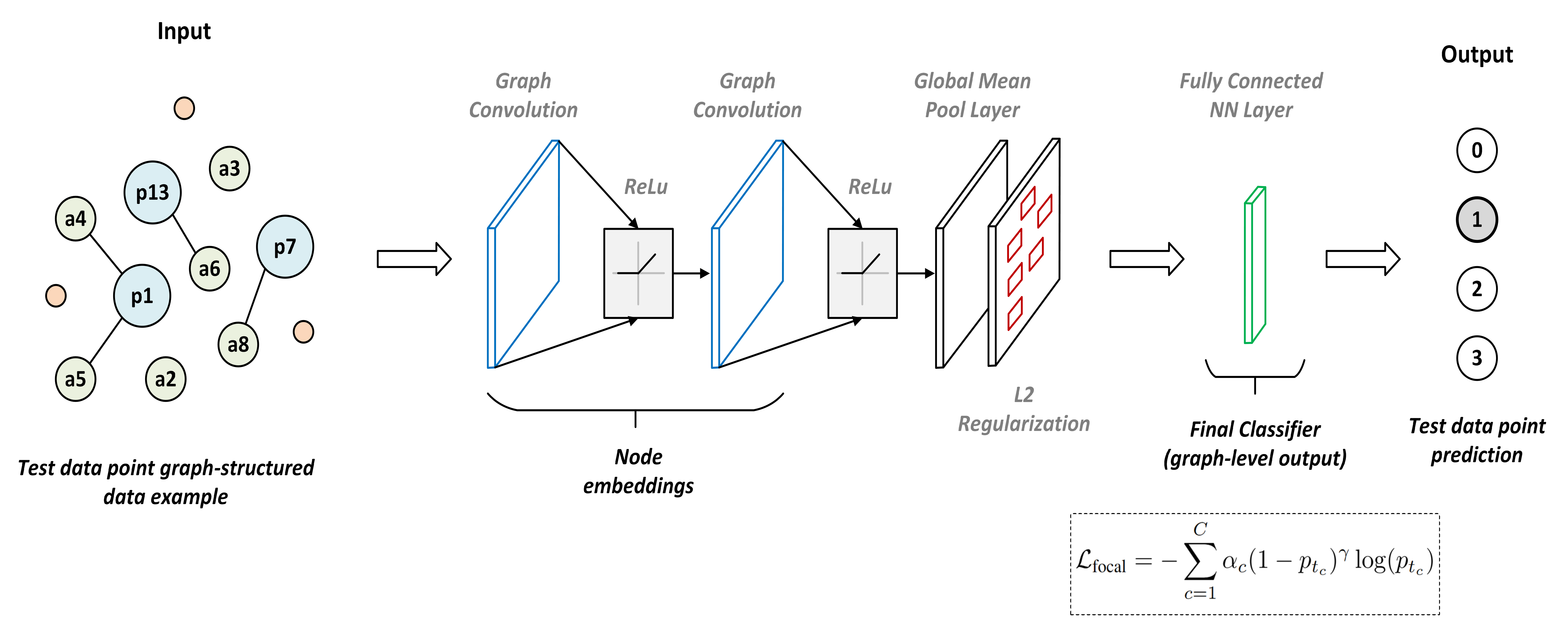}
    \end{adjustbox}
\caption{The graph anomaly predictor module.}
\label{fig: GAP Diagram}
\end{figure*}

The GAP module shown in Fig.~\ref{fig: GAP Diagram} consists of a two-layer GCN model and is employed to predict one of the two instances: normal or abnormal (conflict). The output of GAP will consist of one of four labels (0: normal, 1: direct conflict, 2: implicit conflict, and 3: indirect conflict). 

The fully constructed graph structure for the test data point obtained from the GSC module is fed to the GAP module. GCN learns node-level representations from the graph-structured data, which are then aggregated via a global mean pooling into a graph-level representation for final classification. The resultant graph represents the relationships between the applications, the control parameters, and the KPIs. Thereafter, a fully connected neural network (NN) layer is employed to learn the final mapping from the graph-level representations to the target output classes.

The classifier is enhanced by a focal loss function \cite{Focal_loss_2017} to address class imbalance and improve performance on hard-to-classify instances. Integrating focal loss plays a crucial role in conflict management within O-RAN by enabling the model to concentrate on challenging, infrequent, or critical conflict cases, which are rare in real-world scenarios. The focal loss function improves the standard cross-entropy loss by adding two key parameters, alpha and gamma. 

$\alpha_c$ adjusts for class imbalance by balancing the weight between different classes, while $\gamma$ focuses the model on harder examples by reducing the loss for easy, well-classified examples. This helps to prevent the model from being overwhelmed by the majority class in imbalanced datasets and gives more attention to challenging instances. We compute $\alpha_c$ using inverse class weights by calculating the inverse frequency of each class in the dataset. On the other hand, $\gamma$ is manually tuned for model optimization. The focal loss function for multi-class classification is given by:



\begin{equation}
    \mathcal{L}_{\text{focal}} = - \sum_{c=1}^{C} \alpha_c (1 - p_{t_c})^\gamma \log(p_{t_c})
\end{equation}

where:
\begin{itemize}
    \item \( C \) is the number of classes.
    \item \( p_{t_c} \) is the predicted probability for the true class \( c \).
    \item \( \alpha_c \) is the weighting factor for class \( c \), which can help address class imbalance.
    \item \( \gamma \) is the focusing parameter, which down-weights the loss for well-classified examples.
\end{itemize}

For each data point, this loss function is computed by summing the individual losses for each class, where only the true class \( c \) contributes to the loss. \( p_{t_c} \) is defined as the predicted probability for the true class.


GCN operates by propagating node features through the graph, aggregating information from a node’s neighbors, and learning a new representation for each node. The process is typically iterative (layer-wise), with each layer performing a graph convolution operation. 

After passing through two layers of graph convolution, the final node representations are passed through a fully connected layer to perform graph classification. A mean pooling mechanism is applied to the node features. It aggregates information into a graph-level representation before making the final classification. 

Instead of processing the entire graph at once, mini-batch processing is executed. This involves processing smaller subgraphs simultaneously, which helps reduce memory usage and computational time. Furthermore, a stratified K-Fold is implemented to ensure that each fold in the cross-validation process has a similar distribution of classes as the entire dataset during training and validation. This is particularly useful since 
real-world traffic is highly imbalanced. 

Early stopping is implemented in our model to prevent overfitting by monitoring the validation loss during training. The training process is terminated if the model's performance on the validation set does not improve for a specified number of epochs. The delta ensures that only considerable improvements in validation loss are evaluated as indicators of progress, preventing early stopping from being triggered by negligible changes.

The L2 regularization 
is implemented to mitigate overfitting. The L2 regularization adds a penalty to the loss function that prevents the model from giving too much weight to any single feature. This is achieved by squaring the model's weights and adding the result to the loss, causing the model to prefer lower weight values. The equation for L2 regularization is given by:

\begin{equation}
    \mathcal{L}_{\text{L2}} = \lambda \sum_{i=1}^{n} w_i^2,
\end{equation}
where \( \lambda \) is the regularization parameter controlling the strength of the regularization, \( w_i \) is the weight for the \(i\)-th feature in the model, and \( n \) is the total number of weights in the model.

To promote generalization and mitigate overfitting, the model introduces randomness during the dataset splitting process. Specifically, the entire dataset is randomly permuted, which ensures that the training, validation, and test subsets are sampled in a non-deterministic manner at each run. This random shuffling of indices before splitting helps reduce potential biases related to data ordering and promotes robustness across varying data distributions. The detailed GAP pseudocode is shown in Algorithm 3.


\begin{algorithm}
\caption{Graph Structure Creator (GSC)}
\begin{algorithmic}[1]
\renewcommand{\algorithmicrequire}{\textbf{Input:}}
\renewcommand{\algorithmicensure}{\textbf{Output:}}
\REQUIRE Binary-state dataset $\mathcal{B}_t^{(i)} = \{ s_{\mathcal{A}}^{(i)}, s_{\mathcal{P}}^{(i)}, s_{\mathcal{K}}^{(i)} \}, \; i = 1,2,\dots,N$
\ENSURE Graph-structured dataset
\\ \textit{Initialisation} :
\STATE $i \gets 1$
\WHILE{$i \le N$}

    \IF{$s_{a_j}^{(i)} = 1$ \textbf{ and } $s_{p_j}^{(i)} = 1$}
        \STATE Create subgraph $G^{AP}$ connecting xApps and parameters
    \ELSE
        \STATE No subgraph created
    \ENDIF

    \IF{$s_{p_j}^{(i)} = 1$ \textbf{ and } $s_{k_j}^{(i)} = 1$}
        \STATE Create subgraph $G^{KP}$ connecting parameters and KPIs
    \ELSE
        \STATE No subgraph created
    \ENDIF

    \IF{$s_{p'_j}^{(i)} = 1$ \textbf{ and } $s_{p_j}^{(i)} = 1$}
        \STATE Create subgraph $G^{P'P}$ connecting parameters to parameters
    \ELSE
        \STATE No subgraph created
    \ENDIF

    \STATE Merge all created subgraphs ($G^{AP}, G^{KP}, G^{P'P}$) into a single graph $G$

    \STATE $i \gets i + 1$
\ENDWHILE
\RETURN $G$
\end{algorithmic}
\end{algorithm}

\subsection{Root Cause Analyst (RCA)}
\begin{figure*}[t!]
\centering
    \begin{adjustbox}{width=\textwidth}
        \includegraphics[width=\textwidth]{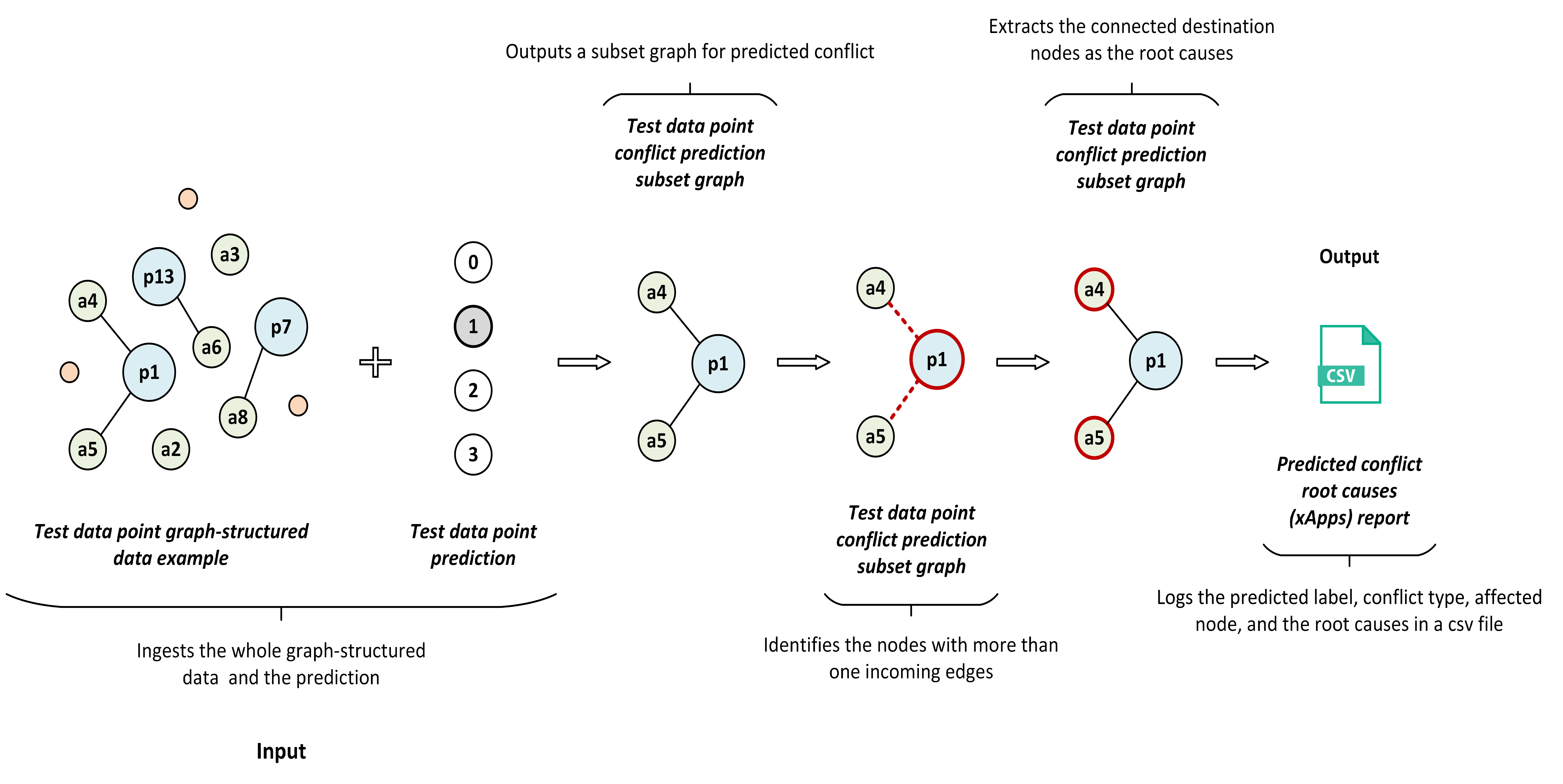}
    \end{adjustbox}
\caption{The root cause analyst (RCA) module.}
\label{fig: RCA Diagram}
\end{figure*}

The RCA module offers interpretative insights into the predictions produced by the GAP module by identifying the xApps contributing to the predicted conflicts (see Fig.~\ref{fig: RCA Diagram}). 
The test data point graph-structured data obtained from GAP and the complete graph structure created by the GSC are input into the RCA module. The output comprises a subgraph that delineates the predicted conflicts, explicitly representing the relationships among the elements contributing to these conflicts.

The subset of the graph corresponding to each prediction is further analyzed. RCA checks the source node with more than one incoming edge, as shown in Fig.~\ref{fig: conflicts_examples}. This means that the behavior of this node is being altered by more than one node. Then, RCA extracts the connected destination nodes as the root causes contributing to the predicted conflict. 

Addressing the root causes proactively improves long-term performance and system stability. A proactive approach relies on a deep understanding of the factors contributing to a problem. By analyzing data to uncover the root causes of service degradation in the network, mobile operators can make informed decisions to avoid reactive responses that are often incomplete or insufficient. 
    



\section{Simulation Setup and Results}
\label{sec:experiments_and_results}
This section presents the experimental setup and performance of the proposed solution. The experimental setup outlines the environment, tools, and methodologies employed to assess the effectiveness of GRAPHICA, while the results provide a comprehensive interpretation and analysis of its performance.

\begin{figure}[ht!]
\centering
    \begin{adjustbox}{width=0.5\textwidth}
        \includegraphics[width=0.5\textwidth]{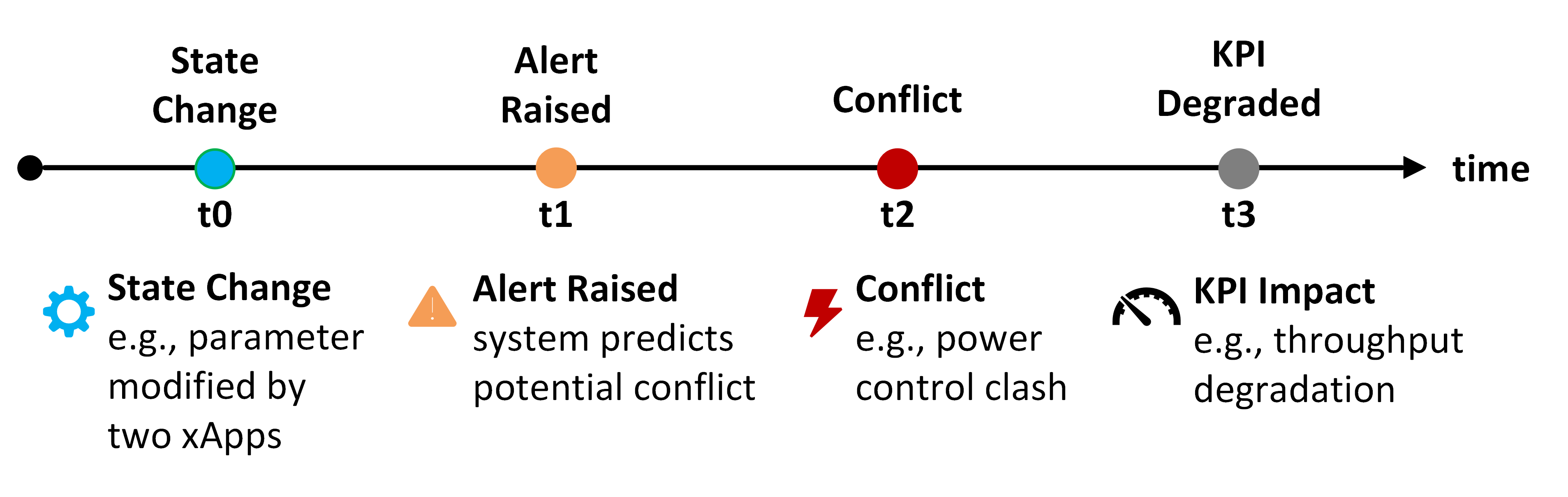}
    \end{adjustbox}
\caption{The chronological sequence of events for conflict prediction.}
\label{fig: GRAPHICA_timeline}
\end{figure}

The chronological sequence of events is illustrated in Fig.~\ref{fig: GRAPHICA_timeline}, highlighting the progression from system monitoring to conflict occurrence. The predictive model utilizes real-time data and a GCN-based model to identify key triggers based on dynamic variables in the network and forecast the likelihood of a conflict occurrence.

Specifically, at time t0, the proposed system initiates the monitoring process and collects data on the state dynamic behavior, including the xApps subscription, operational parameters, and KPI values. At time t1, an alert is raised based on the observed patterns if there is a likelihood of a deviation in the system's behavior. Then, the possible root causes (xApps) are identified to trigger a need for immediate attention and intervention. 
If the alert is raised promptly before the onset of the conflict at time t2, we can avoid a later KPI degradation at time t3. In the remainder of this section, we will describe each module in the proposed method, outlining the corresponding steps and functionality. 

\subsection{Simulation Setup}
\begin{figure*}[t!]
\centering
    \begin{adjustbox}{width=0.7\textwidth}
        \includegraphics[width=\textwidth]{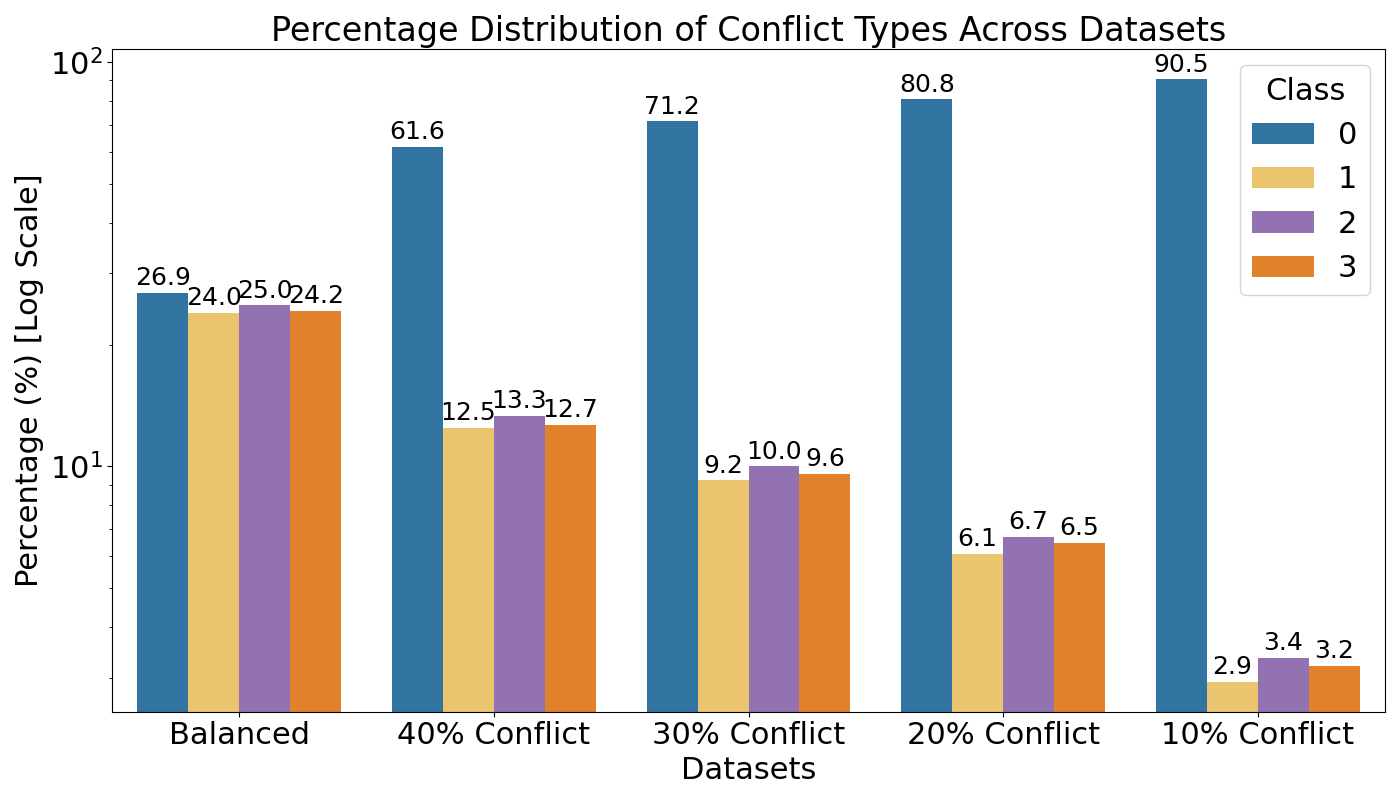}
    \end{adjustbox}
\caption{The label distribution across the binary-state KPIs datasets.}
\label{fig: datasets_distribution}
\end{figure*}

The evaluation was conducted on five datasets exhibiting varying degrees of class balance, as shown in Fig.~\ref{fig: datasets_distribution}. Each dataset contains a mixture of normal and conflict-labeled instances categorized into direct, implicit, and indirect conflict types. The first dataset is balanced, where the classes are evenly distributed with 671 samples per class and a total of 2,684 samples.

The remaining four datasets retain a high number of samples for the normal class as the majority class, while the minority classes represent varying proportions of conflicts, ranging from 40\% to 10\%; the specific number of samples per class can be found in Fig.~\ref{fig: datasets_distribution}. The ``40\% Conflict" dataset consists of a total of 1,145 instances, comprising 689 normal samples, 145 direct conflicts, 158 implicit conflicts, and 153 indirect conflicts. The ``30\% Conflict" dataset includes 977 instances, with 683 labeled as normal, 92 as direct conflicts, 102 as implicit conflicts, and 100 as indirect conflicts. The ``20\% Conflict" dataset contains 848 instances, comprising 678 normal samples, 53 direct conflicts, 59 implicit conflicts, and 58 indirect conflicts. The ``10\% Conflict" dataset presents 752 instances, with 673 normal samples, 25 direct conflicts, 27 implicit conflicts, and 27 indirect conflicts. While distribution among the minorities can vary, we begin with the chosen cases in our study as a starting point for investigation due to the lack of prior knowledge.

The variations in the proportion of conflict instances across the datasets are intended to model real-world scenarios where conflicts are infrequent. This allows us to evaluate the model's performance across datasets with different levels of conflict prevalence. The core objective of this approach is to assess the model's robustness and effectiveness in handling data distributions that closely resemble actual network conditions, where conflict instances are relatively rare.

To evaluate the performance of our method, we use precision (Prec), recall (Rec) and F1-Score (F1) over the test dataset and its ground truth values: $\mathrm{F} 1=\frac{2 \times \operatorname{Prec} \times \text { Rec }}{\text { Prec }+ \text { Rec }}$, where $\operatorname{Prec}=\frac{\mathrm{TP}}{\mathrm{TP}+\mathrm{FP}}$ and $\operatorname{Rec}=\frac{\mathrm{TP}}{\mathrm{TP}+\mathrm{FN}}$. TP, TN, FP, and FN represent true positives, true negatives, false positives, and false negatives, respectively. Our generated datasets are highly imbalanced, which justifies the choice of these metrics that are suitable for imbalanced data. 


We simulated and tested our model on an NVIDIA DGX station Intel(R) Xeon(R) CPU ES-2698 v4 @2.20GHz, 20core/40ht with 256GB of memory. 
The model is trained using the Adam optimizer with a learning rate of 0.01, ReLU activation function, and a weight decay of 1e-4. The training process utilizes a batch size of 128, employing 5-fold cross-validation over 2,000 epochs. A two-layer GCN model operates on the graph's node features. The model performs graph convolutions and aggregates information from the graph structure. 
Global mean pooling \cite{gobal_mean_pooling} is applied to return batch-wise, graph-level outputs by averaging node features across the node dimension.




\subsection{Simulation Results}
\begin{figure}[t!]
    \centering   
    \begin{adjustbox}{width=0.5\textwidth}
        \includegraphics[trim=0cm 0cm 0cm 0cm, width=\textwidth]{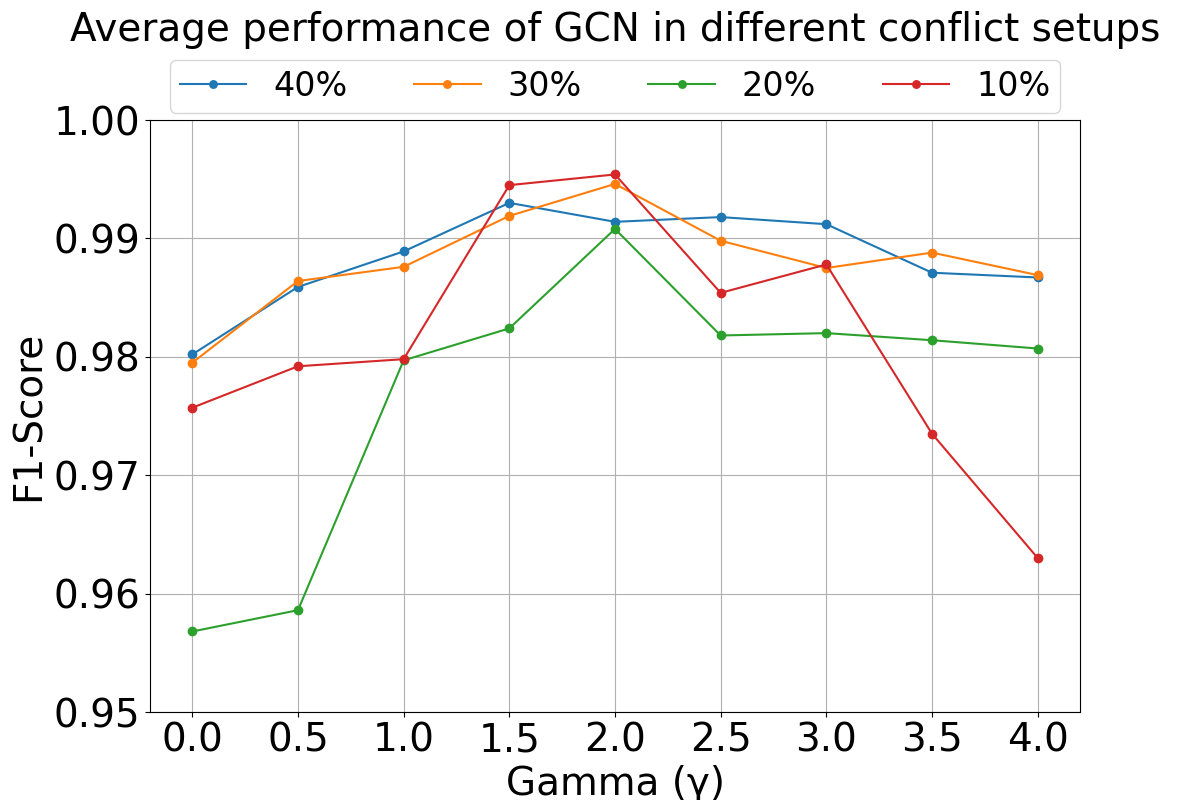}
    \end{adjustbox}
    \caption{Average performance of GCN in different setups.}
    \label{fig: average_performance_of_GCN_in_different_setups}
\end{figure}

We conducted a parametric study provided by Table~\ref{table: GCN Testing Evaluation metrics} and Fig. \ref{fig: average_performance_of_GCN_in_different_setups} to demonstrate the model's performance under different values of  $\gamma$, where the imbalance may have a strong impact on the model's effectiveness.

For the balanced dataset, $\gamma$ is set to zero in the focal loss function as the classes are equally distributed. All classes contribute similarly to the loss function, making the model less likely to be biased toward the majority class. 
The model demonstrates remarkable performance on the balanced dataset, achieving high precision, recall, and F1 of 0.9829, 0.9818, and 0.9817, respectively.


The performance of the proposed model was evaluated for $\gamma$ values ranging from 0.0 to 4.0, with a step size of 0.5, and the average performance was computed across ten independent experiments.
This analysis was performed on the imbalanced datasets under investigation, as shown in Table~\ref{table: GCN Testing Evaluation metrics}. As $\gamma$ increases, the performance metrics improve for all the datasets. The best performance is attained when $\gamma=1.5$ for the 40\% conflict dataset and $\gamma=2.0$ for the 30\%, 20\%, and 10\% conflict datasets; the corresponding F1-scores are 0.9930, 0.9946, 0.9908, and 0.9954, respectively. A greater $\gamma$ leads to degraded performance; this is expected as a larger $\gamma$ causes the model to overly focus on misclassified examples, potentially leading to overfitting. This, in turn, reduces the model’s ability to generalize to easier examples, resulting in a decrease in overall performance.  


The variation of $\gamma$ significantly influences the model's performance. As $\gamma$ increases, the model emphasizes hard-to-classify examples during the training phase, thereby enhancing recall by improving its ability to identify difficult or underrepresented cases. However, this focus on challenging examples can reduce precision, as the model may misclassify some simpler instances. Conversely, when $\gamma$ is set to a lower value, the model treats all examples with equal importance during training, resulting in a more balanced performance. This may limit the model's capacity to focus on difficult cases adequately. Therefore, selecting the optimal $\gamma$ value is crucial for achieving a balance between precision and recall, thereby ensuring the model performs effectively without overemphasizing any particular type of conflict. Overall, as shown in Table~\ref{table: GCN Testing Evaluation metrics}, the model exhibits consistent precision, recall, and F1 across different test datasets with varying levels of imbalance.

\begin{figure}[t!]
    \centering   
    \begin{adjustbox}{width=0.5\textwidth}
        \includegraphics[trim=0cm 0cm 0cm 0cm, width=0.7\textwidth]{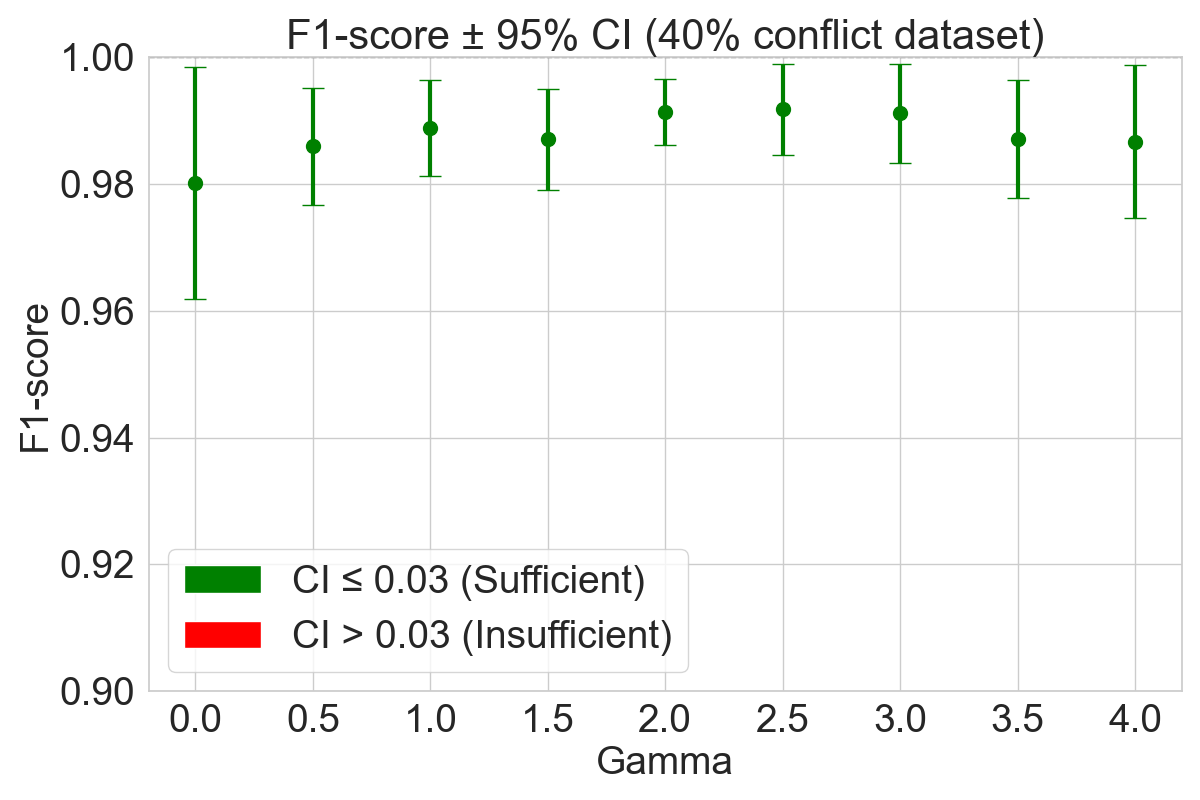}
    \end{adjustbox}
    \caption{Average F1 scores and their confidence intervals across multiple runs (40\% conflict dataset).}
    \label{fig: confidence intervals_40_pt_conflict_dataset}
\end{figure}

\begin{figure}[t!]
    \centering   
    \begin{adjustbox}{width=0.5\textwidth}
        \includegraphics[trim=0cm 0cm 0cm 0cm, width=0.7\textwidth]{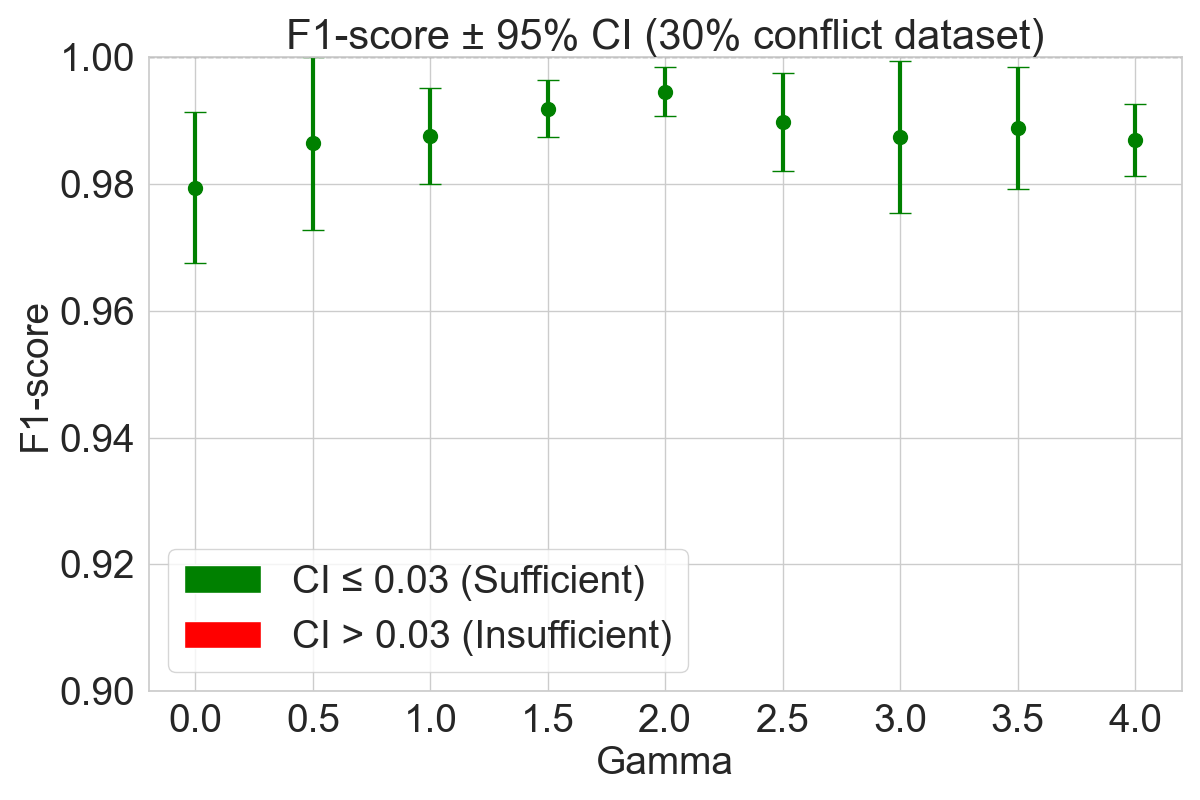}
    \end{adjustbox}
    \caption{Average F1 scores and their confidence intervals across multiple runs (30\% conflict dataset).}
    \label{fig: confidence intervals_30_pt_conflict_dataset}
\end{figure}

\begin{figure}[t!]
    \centering   
    \begin{adjustbox}{width=0.5\textwidth}
        \includegraphics[trim=0cm 0cm 0cm 0cm, width=0.7\textwidth]{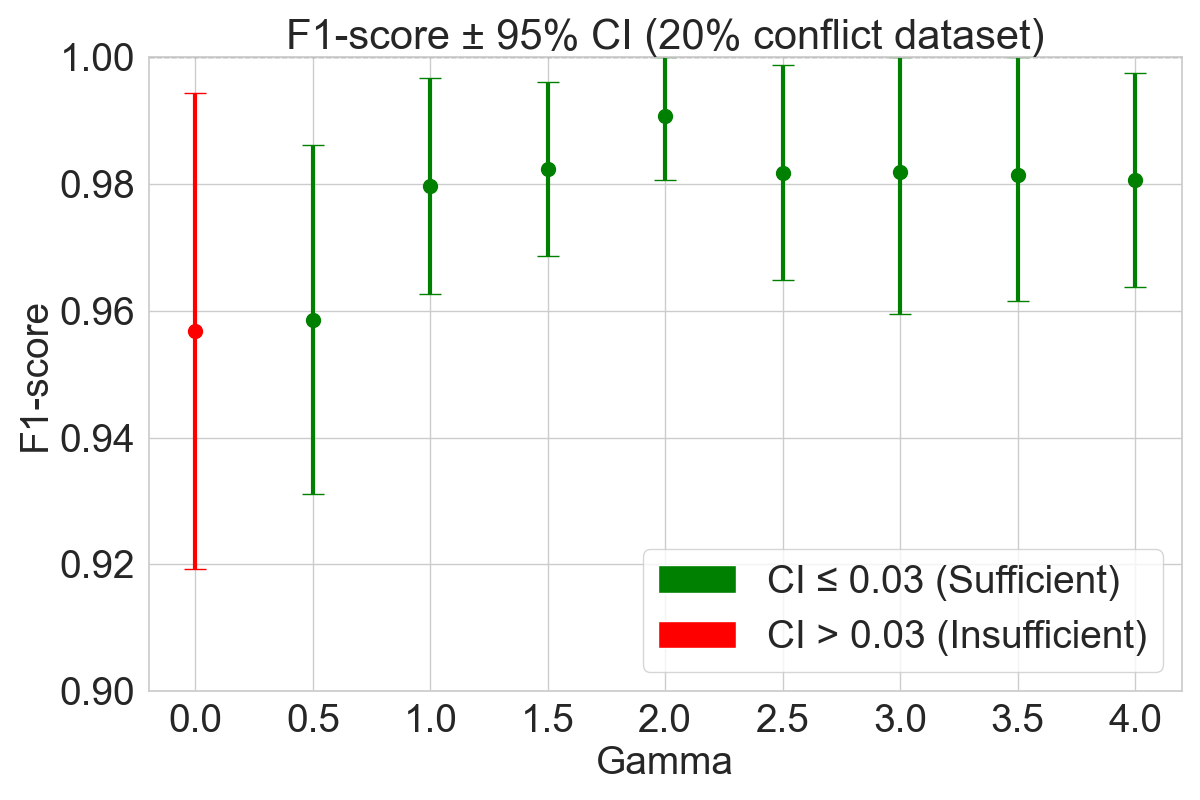}
    \end{adjustbox}
    \caption{Average F1 scores and their confidence intervals across multiple runs (20\% conflict dataset).}
    \label{fig: confidence intervals_20_pt_conflict_dataset}
\end{figure}

\begin{figure}[t!]
    \centering   
    \begin{adjustbox}{width=0.5\textwidth}
        \includegraphics[trim=0cm 0cm 0cm 0cm, width=0.7\textwidth]{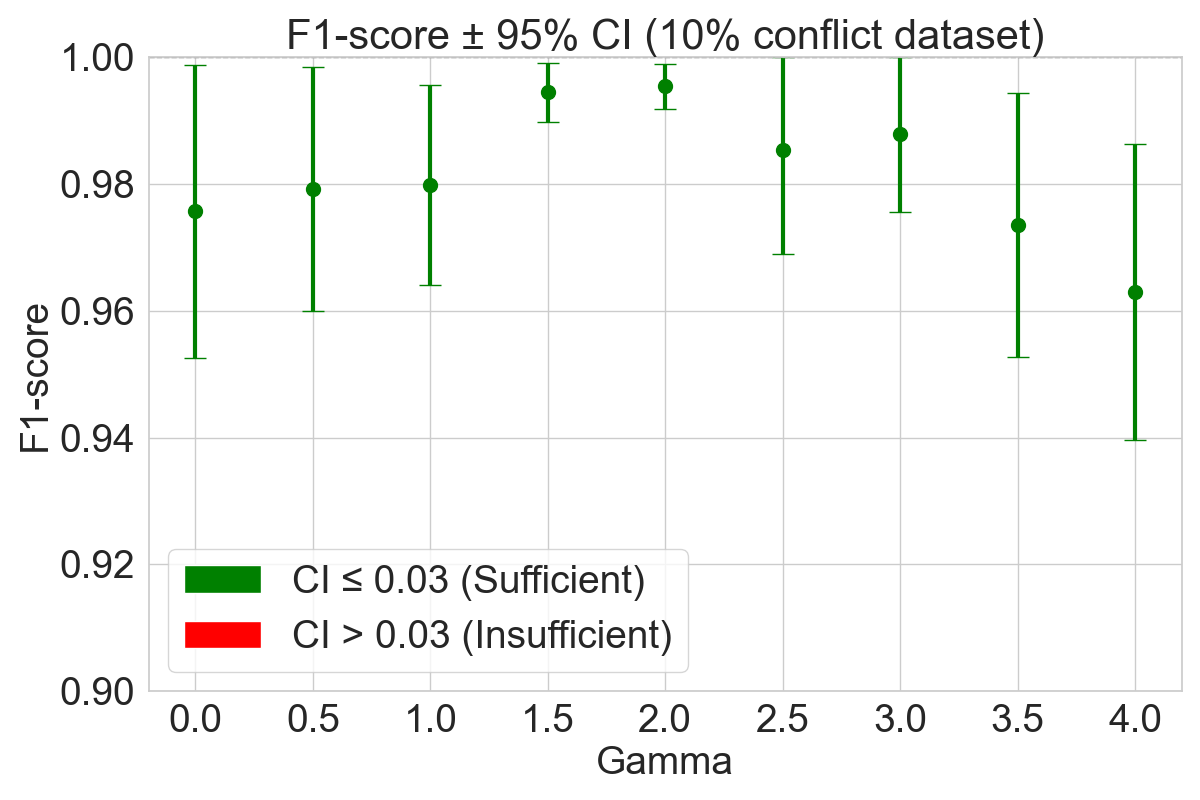}
    \end{adjustbox}
    \caption{Average F1 scores and their confidence intervals across multiple runs (10\% conflict dataset).}
    \label{fig: confidence intervals_10_pt_conflict_dataset}
\end{figure}

The average F1 scores with their corresponding 95\% confidence intervals across multiple runs and gamma values for each dataset are shown in Fig.~\ref{fig: confidence intervals_40_pt_conflict_dataset}--\ref{fig: confidence intervals_10_pt_conflict_dataset}. In our evaluation, the number of samples is calculated as the product of the number of experimental runs and the number of independent performance scores obtained per run. Specifically, we conducted 10 runs, each employing 5-fold cross-validation, resulting in a total of 10×5=50 independent samples. These plots are essential for assessing the reliability and consistency of the experimental results. By averaging the results over 10 independent runs for each setting, we can estimate the variability in model performance and determine whether the observed trends are statistically significant. The 95\% confidence intervals, computed using Student’s t-distribution \cite{Fuzzy_STUDENT’S_T-Distribution_Model_2022}, indicate the range within which the true F1 score is expected to fall with high certainty. Most intervals are narrow, suggesting stable performance and supporting the conclusion that 10 runs are sufficient for reliable evaluation. To enhance interpretability, we included the number of runs on each point and color-coded the error bars: red bars signal wider intervals (greater than ±0.03). This implies that more runs may be needed for those specific settings. Overall, these visualizations offer a robust framework for evaluating performance stability and verifying the adequacy of our experimental design.

\begin{table*}[t!]
\centering
\caption{Average Performance of GCN in different setups for 10 iterations for the binary-state KPIs test datasets.}
\begin{adjustbox}{width=\textwidth}
\begin{tabular}{|c|c|c|c|c|c|c|c|c|c|c|}                
\hline
\multicolumn{1}{|c|}{\multirow{2}{*}{\textbf{Conflict (\%)}}} & \multicolumn{10}{c|}{\textbf{Performance}}                                                                                           \\ \cline{2-11} 
\multicolumn{1}{|c|}{}            &  \textbf{Metric}    & $\gamma$=0.0 & $\gamma$=0.5 & $\gamma$=1.0 & $\gamma$=1.5  & $\gamma$=2.0     & $\gamma$=2.5 & $\gamma$=3.0    & $\gamma$=3.5 & $\gamma$=4.0                 \\ \hline

\multirow{3}{*}{40}    & Prec      & 0.9849     & 0.9879     & 0.9900     & 0.9937            & 0.9923            & 0.9930     & 0.9923            & 0.9889     & 0.9892                     \\ \cline{2-11} 
                                  & Rec       & 0.9809     & 0.9861     & 0.9890     & 0.9931            & 0.9913            & 0.9919     & 0.9913            & 0.9873     & 0.9867                     \\ \cline{2-11} 
                                  & F1        & 0.9802     & 0.9859     & 0.9889     & \textbf{0.9930}   & 0.9914            & 0.9918     & 0.9912            & 0.9871     & 0.9867                     \\ \hline
\multirow{3}{*}{30}    & Prec      & 0.9837     & 0.9884     & 0.9897     & 0.9929            & 0.9951            & 0.9915     & 0.9904            & 0.9906     & 0.9886                     \\ \cline{2-11} 
                                  & Rec       & 0.9804     & 0.9865     & 0.9878     & 0.9919            & 0.9946            & 0.9899     & 0.9878            & 0.9892     & 0.9872                     \\ \cline{2-11} 
                                  & F1        & 0.9795     & 0.9864     & 0.9876     & 0.9919            & \textbf{0.9946}   & 0.9898     & 0.9875            & 0.9888     & 0.9869                     \\ \hline
\multirow{3}{*}{20}    & Prec      & 0.9648     & 0.9633     & 0.9792     & 0.9848            & 0.9939            & 0.9854     & 0.9834            & 0.9885     & 0.9867                     \\ \cline{2-11} 
                                  & Rec       & 0.9633     & 0.9641     & 0.9828     & 0.9836            & 0.9914            & 0.9820     & 0.9836            & 0.9828     & 0.9820                     \\ \cline{2-11} 
                                  & F1        & 0.9568     & 0.9586     & 0.9797     & 0.9824            & \textbf{0.9908}   & 0.9818     & 0.9820            & 0.9814     & 0.9807                     \\ \hline
\multirow{3}{*}{10}    & Prec      & 0.9721     & 0.9760     & 0.9913     & 0.9954            & 0.9960            & 0.9871     & 0.9894            & 0.9766     & 0.9605                     \\ \cline{2-11} 
                                  & Rec       & 0.9816     & 0.9842     & 0.9798     & 0.9947            & 0.9956            & 0.9860     & 0.9895            & 0.9781     & 0.9702                     \\ \cline{2-11} 
                                  & F1        & 0.9757     & 0.9792     & 0.9798     & 0.9945            & \textbf{0.9954}   & 0.9854     & 0.9878            & 0.9735     & 0.9630                     \\ \hline
\end{tabular}
\label{table: GCN Testing Evaluation metrics}

\end{adjustbox}
\end{table*}

\begin{figure}[t!]
    \centering   
    \begin{adjustbox}{width=0.5\textwidth}
        \includegraphics[trim=0cm 0cm 0cm 0cm, width=0.7\textwidth]{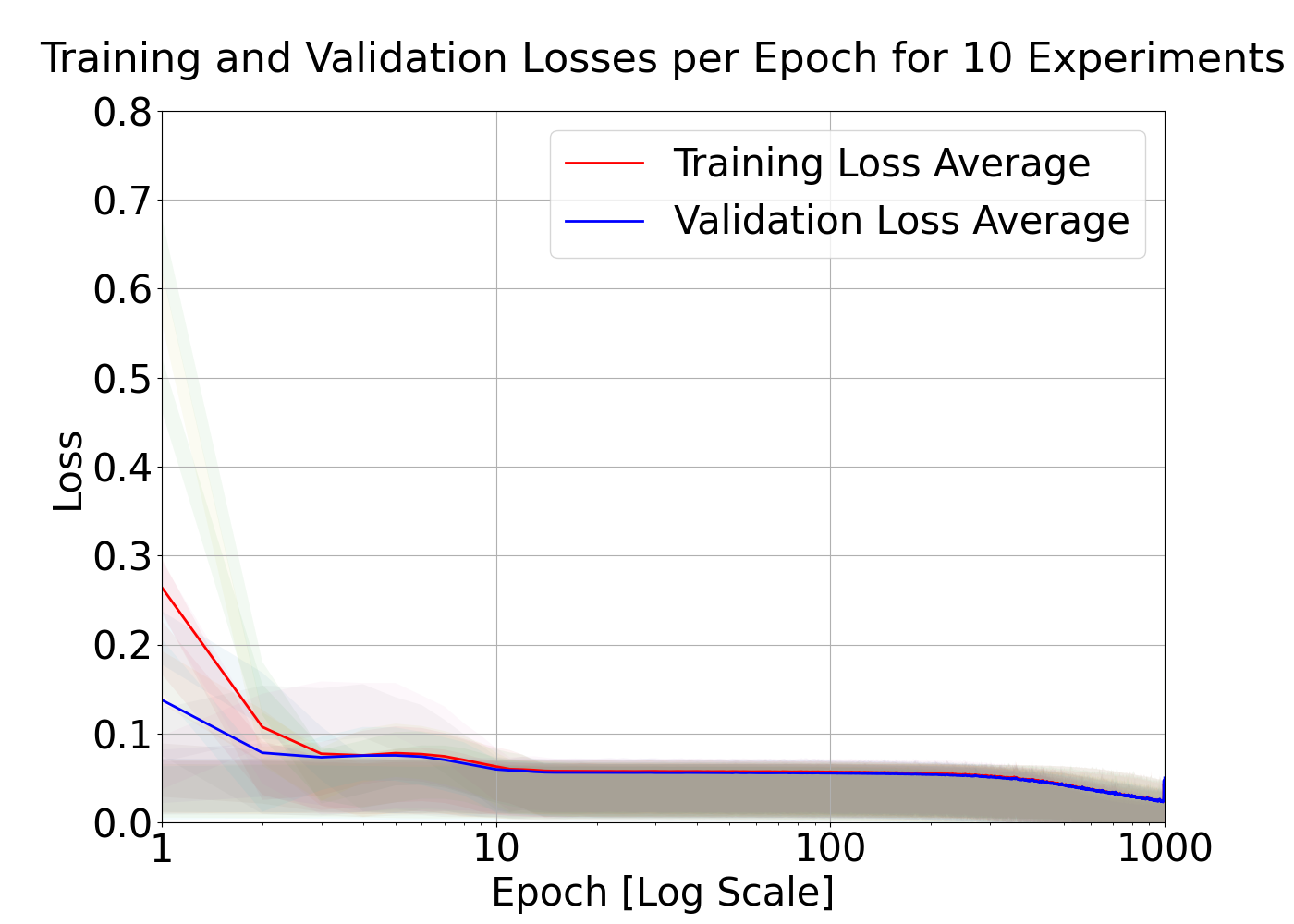}
    \end{adjustbox}
    \caption{Training and validation losses per epoch on the 10\% conflict dataset.}
    \label{fig: 90_normal_dataset_GNN_training_validation_loss}
\end{figure}

For further illustration, the training and validation losses across epochs for 10 experiments for the 10\% conflict dataset ($\gamma$ = 2.0) are shown in Fig.~\ref{fig: 90_normal_dataset_GNN_training_validation_loss}, represented as shaded regions to show variability, with bold lines indicating the average training and validation trends. The losses decrease throughout training. This indicates that the model is effectively learning from the data over time. Initially, both losses were high. This reflects the model's early stage of learning, where it struggles to fit the data. As the training progresses, the loss decreases, indicating that the model is gradually learning to make better predictions on the training data. Thus, the model improves its ability to fit the training data and gradually minimizes the error between its predictions and the actual values. Similarly, the validation loss decreases, suggesting that the model is generalizing well and not overfitting to the training set. Both losses decrease at a similar rate and converge to a low value of less than 0.05. The decreasing losses typically point to a well-optimized model, learning meaningful patterns and avoiding excessive complexity or bias. 

\begin{figure}[t!]
    \centering   
    \begin{adjustbox}{width=0.4\textwidth}
        \includegraphics[trim=0cm 0.2cm 0cm 0cm, width=1\textwidth]{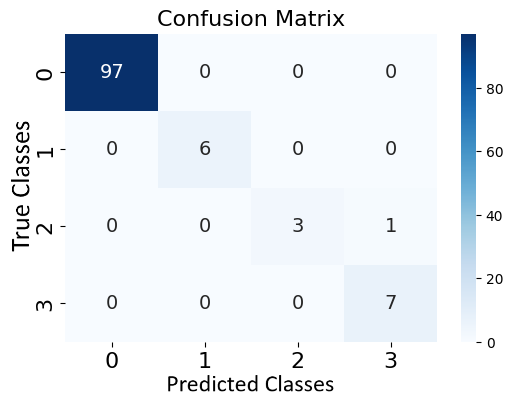}
    \end{adjustbox}
    \caption{Confusion matrix on the 10\% conflict test dataset.}
    \label{fig: 90_normal_dataset_GNN_confusion_matrix}
\end{figure}

The confusion matrix (CM) of the proposed method testing on the 10\% conflict dataset ($\gamma$ = 2.0) is shown 
in Fig.~\ref{fig: 90_normal_dataset_GNN_confusion_matrix}. 
The test set consists of 114 data points. Among these, 97 are predicted as normal, 6 as direct conflicts, 3 as implicit conflicts, and 7 as 
indirect conflicts. In this highly imbalanced scenario, where normal cases significantly outnumber conflicts, only one implicit conflict was misclassified as an indirect.
This suggests that the model is capable of generalizing effectively on imbalanced datasets.



\begin{table}[t!]
\caption{An example of conflict reports by the RCA (10\% conflict)}
\begin{adjustbox}{width=0.49\textwidth}
\begin{tabular}{|>{\centering\arraybackslash}p{1.0cm}|>{\centering\arraybackslash}p{1cm}|>{\centering\arraybackslash}p{1.0cm}|>{\centering\arraybackslash}p{1.55cm}|>{\centering\arraybackslash}p{1.55cm}|}
\hline
\textbf{Predicted Label} & \textbf{Conflict Type} & \textbf{Affected Node} & \textbf{Root Causes Nodes} & \textbf{Root Causes xApps} \\ \hline        
1                       & Direct               & p12                     & a1, a3               & a1, a3                 \\ \hline
3                       & Indirect             & p8                      & p1, p3               & a1, a3                 \\ \hline
1                       & Direct               & p2                      & a4, a6               & a4, a6                 \\ \hline
2                       & Implicit             & k10                     & p1, p9               & a1, a9                 \\ \hline
1                       & Direct               & p10                     & a2, a5, a8           & a2, a5, a8             \\ \hline
3                       & Indirect             & p13                     & p7, p9, p10          & a7, a9, a10            \\ \hline
2                       & Implicit             & k5                      & p4, p2               & a4, a2                 \\ \hline

\end{tabular}
\label{table: 10_pct_conflict_RCA_report_3.5}
\end{adjustbox}
\end{table}

The conflict RCA report for the 10\% conflict test dataset ($\gamma$ = 2.0) is illustrated in Table~\ref{table: 10_pct_conflict_RCA_report_3.5}. The normal data points in the table are filtered to emphasize the predicted conflicts. The report elucidates the predicted label, the type of conflict, the affected node, and the root causes (nodes and xApps). The affected node is the one whose behavior is altered by one or more nodes. The root cause nodes are the nodes that are concurrently modifying the same node, and the root causes (xApps) are the applications contributing to the predicted conflict. 

As an example, the first line of the report indicates the occurrence of a direct conflict (predicted label = 1). The value of the parameter p12 is influenced by the xApps a1 and a3. Consequently, the root causes of this direct conflict are identified as the xApps a1 and a3. The last line of the report reveals an implicit conflict, with the predicted label being 2. In particular, the value of kpi k5 is affected by two parameters (p4 and p2), which are controlled by the xApps a4 and a2, respectively. Therefore, the root causes of this conflict are the xApps a4 and a2. 

We benchmark our proposed method using the dataset employed in \cite{Learning_and_Reconstructing_Conflicts_in_O-RAN_2024} to ensure a fair and consistent comparison with the GraphSAGE method, a graph-based learning approach that shares methodological similarities with our work. This evaluation framework enables us to assess the relative performance of our approach under identical data conditions, thereby providing a direct and reliable comparison between the two methods. In this dataset, the KPIs are modeled to follow a Gaussian distribution, with their values determined by equations that characterize the relationships between control parameters and KPIs, as detailed below:

\begin{align}
k_1 &= 0.5 \cdot \exp\left(-\frac{(p_1 + 50)^2}{(2 p_2)^2}\right) \\
k_2 &= \exp\left(-\frac{(p_1 - 50)^2}{(2 p_3)^2}\right) \\
k_3 &= \exp\left(-\frac{(p_4 + k_1)^2}{(2 p_5)^2}\right) \\
k_4 &= \exp\left(-\frac{(p_7 + k_2)^2}{(2 p_6)^2}\right)
\end{align}

\begin{figure*}[t!]
\centering
    \begin{adjustbox}{width=0.7\textwidth}
        \includegraphics[width=\textwidth]{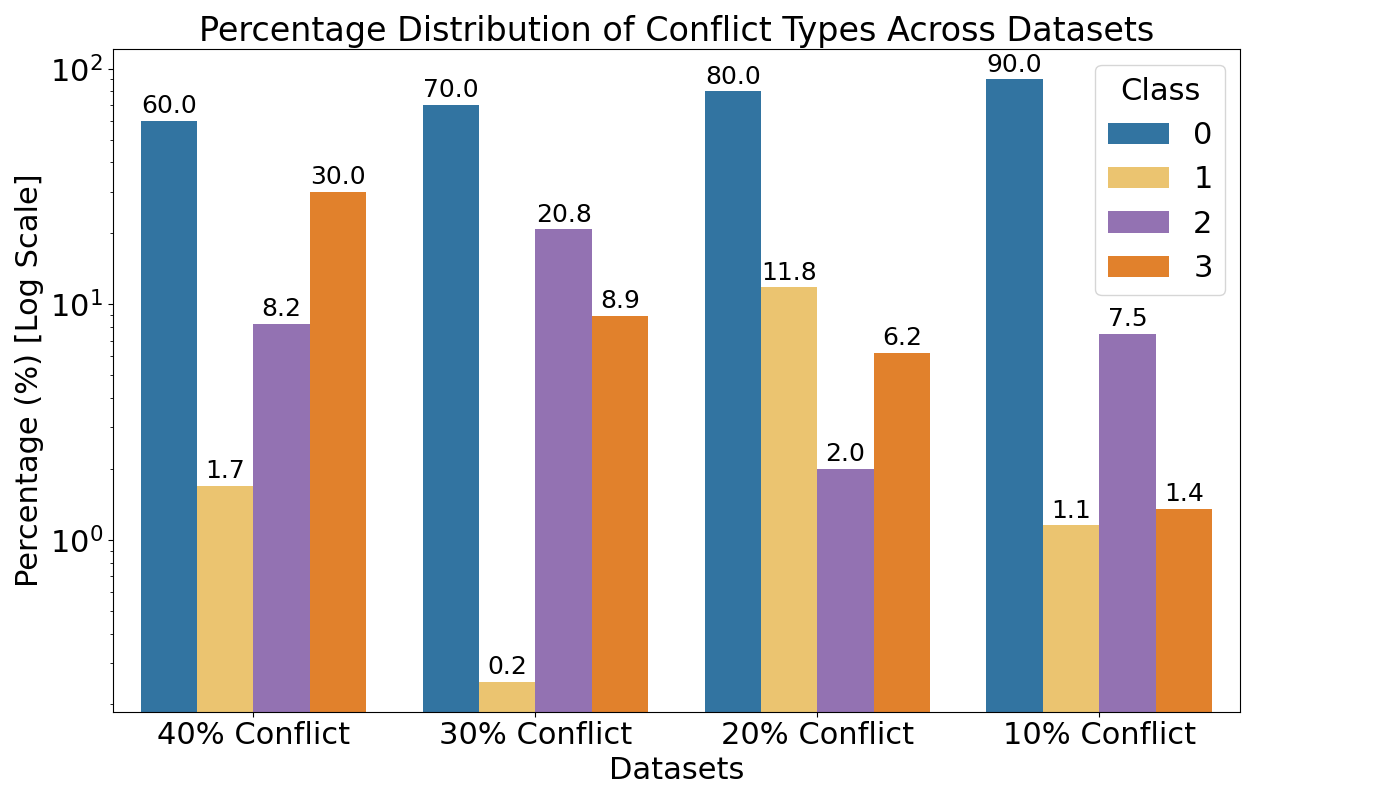}
    \end{adjustbox}
\caption{The label distribution across the Gaussian distribution KPIs datasets.}
\label{fig: datasets_distribution_gaussian_kpis}
\end{figure*}

Based on this model, the control parameters are sampled using uniformly distributed random values within predefined ranges \cite{Learning_and_Reconstructing_Conflicts_in_O-RAN_2024}. 
We generated four highly imbalanced datasets with varying proportions of conflict (40\%, 30\%, 20\%, and 10\%) as shown in Fig.~\ref{fig: datasets_distribution_gaussian_kpis} and evaluated the model’s performance across ten independent experiments. The corresponding F1-scores are 0.9812, 0.9965, 0.9951, and 0.9828, respectively. The average performance is summarized in Table~\ref{table: GCN Testing Evaluation metrics for Gaussian distribution KPIs}. 
These findings indicate that the model reliably achieves elevated F1-scores across all datasets, thereby evidencing substantial performance across diverse conflict scenarios.

\begin{table*}[t!]
\centering
\caption{Average Performance of GCN in different setups for 10 iterations for the Gaussian distribution KPIs test datasets ($\gamma=2.0$).}
\label{table: GCN Testing Evaluation metrics for Gaussian distribution KPIs}
\begin{adjustbox}{width=0.6\textwidth}
\begin{tabular}{|c|c||c|c|}
\hline
\textbf{Conflict (\%)} & \textbf{Performance} & \textbf{Conflict (\%)} & \textbf{Performance} \\ \hline

40 & 
\begin{tabular}{@{}l r@{}}
\textbf{Metric} & \textbf{Value} \\ \hline
Precision & 0.9826 \\ \hline
Recall    & 0.9827 \\ \hline
F1-score  & 0.9812
\end{tabular} &

20 & 
\begin{tabular}{@{}l r@{}}
\textbf{Metric} & \textbf{Value} \\ \hline
Precision & 0.9958 \\ \hline
Recall    & 0.9957 \\ \hline
F1-score  & 0.9951
\end{tabular} \\ \hline

30 & 
\begin{tabular}{@{}l r@{}}
\textbf{Metric} & \textbf{Value} \\ \hline
Precision & 0.9954 \\ \hline
Recall    & 0.9977 \\ \hline
F1-score  & 0.9965
\end{tabular} &

10 & 
\begin{tabular}{@{}l r@{}}
\textbf{Metric} & \textbf{Value} \\ \hline
Precision & 0.9785 \\ \hline
Recall    & 0.9880 \\ \hline
F1-score  & 0.9828
\end{tabular} \\ \hline

\end{tabular}
\end{adjustbox}
\end{table*}

\begin{algorithm}[t]
\caption{Graph Anomaly Predictor (GAP)}
\begin{algorithmic}[1]
\renewcommand{\algorithmicrequire}{\textbf{Input:}}
\renewcommand{\algorithmicensure}{\textbf{Output:}}
\REQUIRE Graph dataset $\mathcal{G}$, learning rate $\eta$, regularization $\lambda$, focal loss parameters $(\alpha_c, \gamma)$, patience $P$, delta $\delta$, maximum epochs $max\_epochs$
\ENSURE Trained model $f_\theta$ with optimal parameters $\theta^*$ and predicted labels $\hat{y}$

\STATE Shuffle dataset indices
\STATE Split dataset $\mathcal{G}_\pi$ into training, validation, and test subsets
\STATE Apply stratified $K$-fold partitioning to preserve class distribution

\STATE Initialize node features $X \in \mathbb{R}^{|\mathcal{V}| \times d}$
\FOR{each GCN layer $l = 1, \dots, L$}
    \STATE $H^{(l)} \gets \sigma \big( \hat{D}^{-1/2} \hat{A} \hat{D}^{-1/2} H^{(l-1)} W^{(l)} \big)$ 
    \STATE where $\hat{A} = A + I$ and $\hat{D}$ is the degree matrix
\ENDFOR
\STATE Obtain final node embeddings $H^{(L)}$

\STATE Apply mean pooling: $h_\mathcal{G} \gets \frac{1}{|\mathcal{V}|} \sum_{v \in \mathcal{V}} H^{(L)}_v$
\STATE Compute logits via fully connected layer: $z \gets W_{fc} h_\mathcal{G} + b$

\STATE Compute focal loss for true class $c$: $\mathcal{L}_{FL} \gets - \alpha_c (1 - p_{t_c})^\gamma \log(p_{t_c})$, $p_{t_c} = \text{softmax}(z)_c$
\STATE Add L2 regularization: $\mathcal{L}_{total} \gets \mathcal{L}_{FL} + \lambda \sum_i ||w_i||_2^2$

\STATE Initialize optimizer (Adam) with learning rate $\eta$
\STATE Set $epoch \gets 0$, $no\_improve \gets 0$, $\mathcal{L}_{val}^{best} \gets \infty$
\WHILE{$epoch < max\_epochs \; \wedge \; no\_improve < P$}
    \FOR{each mini-batch of subgraphs in training set}
        \STATE Compute forward pass
        \STATE Compute loss $\mathcal{L}_{total}$
        \STATE Backpropagate gradients
        \STATE Update parameters $\theta$ using Adam optimizer
    \ENDFOR
    \STATE Compute validation loss $\mathcal{L}_{val}^{(epoch)}$
    \IF{$\mathcal{L}_{val}^{best} - \mathcal{L}_{val}^{(epoch)} \geq \delta$}
        \STATE $\mathcal{L}_{val}^{best} \gets \mathcal{L}_{val}^{(epoch)}$
        \STATE Save parameters $\theta^*$ corresponding to this epoch
        \STATE $no\_improve \gets 0$
    \ELSE
        \STATE $no\_improve \gets no\_improve + 1$
    \ENDIF
    \STATE $epoch \gets epoch + 1$
\ENDWHILE

\STATE Stop training
\STATE Predict labels: $\hat{y} = \arg\max \text{softmax}(z)$ using saved parameters $\theta^*$
\RETURN $\hat{y}$, $\theta^*$
\end{algorithmic}
\end{algorithm}

\section{Discussions}
\label{sec:qualitative_analysis}
While the demonstrated performance is outstanding on the curated datasets, to the best of our knowledge, there is a lack of publicly available datasets or benchmarks for comparing the performance against other approaches to xApp conflict analysis.
In \cite{QACM_QoS_Aware_xApp_Conflict_Mitigation_2024, A_Game_Theoretic_Approach_2023, Learning_and_Reconstructing_Conflicts_in_O-RAN_2024}, the proposed methods are tested on simulated synthetic data based on the Gaussian distribution function to generate KPIs for xApps. In \cite{xapp_distillation_2024} and \cite{PACIFISTA_2024}, simulations are developed using network emulators such as the Colosseum wireless network emulator and ``mobile-env” wireless communication environment in the Gymnasium framework, respectively. A key extension of the work is to enrich the scenarios and release, where applicable, a benchmark with open data from real or realistic O-RAN in 5G/beyond 5G. This would allow further validation and comparisons of different methods in practical scenarios.

Our results are still sufficiently good 
because the missed conflicts, although mostly implicit, are generally mild. The more severe conflicts are effectively flagged promptly. Our approach is particularly effective in practical applications, as it efficiently enables early prediction of rare conflict scenarios up to 10\%, providing ample time for preventive measures to be put in place.

Existing methods \cite{Team_Learning_Based_Resource_2022,A_Game_Theoretic_Approach_2023,QACM_QoS_Aware_xApp_Conflict_Mitigation_2024,Learning_and_Reconstructing_Conflicts_in_O-RAN_2024} are heavily distribution-dependent, often predicated on the assumption that the underlying data follows a Gaussian distribution, which is used to assign values to KPIs. However, in practical, real-world applications, KPIs frequently exhibit more complex behaviors and may not conform to Gaussian distributions. Our approach addresses this limitation by being distribution-independent, where KPIs are assigned binary state values based on their dynamic behavior. This design enables our methodology to generalize effectively across diverse KPI contexts without being constrained to specific distributions. 

While our proposed approach primarily focuses on conflict prediction and root cause analysis, future work will emphasize the development of a comprehensive predictive maintenance framework that incorporates mitigation strategies based on the identified root causes. This framework aims to proactively prevent performance degradation and reduce xApp conflicts, particularly in complex multi-vendor O-RAN environments. We acknowledge the importance of conflict resolution and plan to integrate state-of-the-art conflict management techniques in subsequent research to complement the prediction capabilities presented here. This will enable a complete conflict management solution, moving beyond prediction toward effective resolution and network optimization.

\section{Conclusion}
\label{sec:conclusion}
In conclusion, this work introduces GRAPHICA, a novel graph-based, event-driven framework designed for conflict prediction and root cause analysis (RCA) in O-RAN environments. The dynamic behavior of xApps, controllable parameters, and Key Performance Indicators (KPIs) is captured using a binary-state dataset, which encodes the variations in their values relative to the previous timestamp. This representation effectively encapsulates the temporal dynamics and operational shifts within the network, forming the basis for constructing graph-structured data that models interdependencies among key network elements. Beyond presenting a new modeling approach, our findings reveal critical insights into the structure and behavior of xApp conflicts, demonstrating that a Graph Convolutional Network (GCN) is highly effective in navigating the complexity of interdependent data. The model’s ability to effectively address the challenges of imbalanced data using a focal loss function and accurately predict direct, indirect, and implicit conflicts reinforces its practical applicability in the real world, where conflicts are rare. Importantly, the RCA module not only localizes the source of conflict but also lays the foundation for actionable mitigation strategies. These contributions mark a significant step toward autonomous and intelligent conflict management in O-RAN systems. Future research will build upon this foundation to investigate proactive coordination mechanisms, conflict prevention policies, multi-conflict prediction techniques, and vendor-agnostic solutions aimed at ensuring reliable, scalable, and efficient operation in next-generation open RAN environments.


%

\bibliographystyle{IEEEtran}
\bibliography{bibliography.bib}
\begin{IEEEbiographynophoto}
{Maryam Al Shami} (Graduate Student Member and Women in Engineering (WIE) Member, IEEE) is pursuing her PhD in Information and Systems Engineering at Concordia University, Canada. Her current research focuses on proactive network management and root-cause analysis in 5G RAN and beyond. Her research interests include 5G networks and beyond, self-healing networks, explainable AI/ML, causal discovery, root cause analysis, and predictive maintenance.
\end{IEEEbiographynophoto}


\begin{IEEEbiography}[{\includegraphics[width=1in,height=1.25in,clip,keepaspectratio]{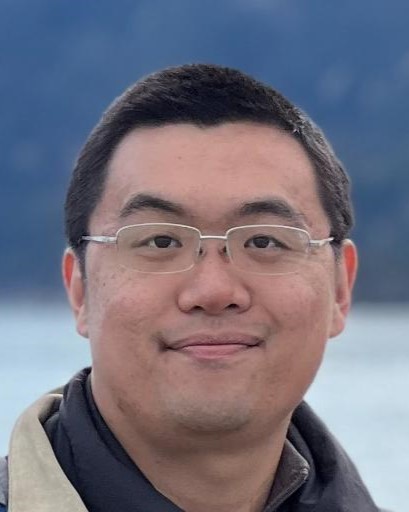}}]{Jun Yan}
    is an Associate Professor and the Research Chair in Artificial Intelligence in Cyber Security and Resilience at Concordia University, Montréal, Canada. He received his B.Eng. in Information and Communication Engineering from Zhejiang University, China, and his Master's and Ph.D. (with Excellence in Doctoral Research) in Electrical Engineering from the University of Rhode Island, USA. He joined Concordia in 2018 and received an early tenure promotion in 2022, where he has served as the Graduate Program Director and Associate Director of the Concordia Institute for Information Systems Engineering. He has co-authored over 100 peer-reviewed research articles and co-founded the Security Research Centre and the Digital Twins Hub at Concordia. His current research interests include computational intelligence, cyber-physical security, and trustworthy AI with applications in smart grids, smart transportation, and smart cities.
\end{IEEEbiography}

\begin{IEEEbiography}[{\includegraphics[width=1in,height=1.25in,clip,keepaspectratio]{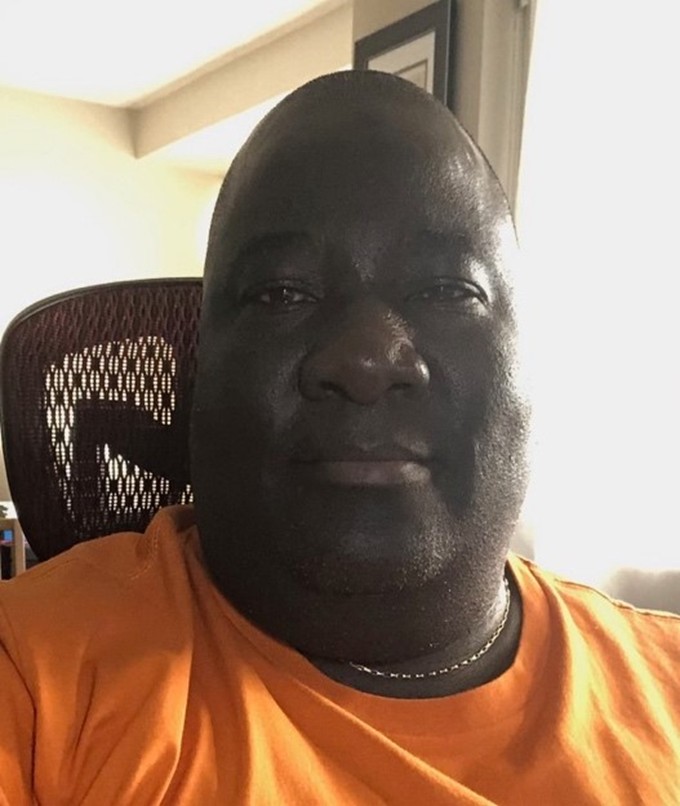}}]{Emmanuel Thepie Fapi} is a Senior Data Scientist at Ericsson Canada, AI Hub Canada, Montreal. Prior to Ericsson, he worked as an audio software developer with Amazon Lab126 (Boston, MA, USA), a software developer/designer with GENBAND U.S. LLC, QNX Software System Limited (Vancouver, BC, CA), an analyst with MDA Systems (Vancouver, BC, CA), and as a DSP engineer with Easy G (Vancouver, BC, CA). He holds a PhD from IMT Atlantique (formerly Ecole Nationale des Télécommunications de Bretagne) in signal processing and telecommunications in Brest, France. He also received a master’s degree in engineering mathematics and computer tools from Orleans University in France. He won the Ericsson Impact Award in 2021 and the Ericsson Key Contributor Award in 2024. He is currently leading the Cognitive AI Applied AI Research pool of projects. He has been involved in 25+ patent applications and 15+ published journal and conference papers. His research interests are 5G Networks and Beyond, Distributed AI/ML, Edge Computing, V2X, Sustainable AI, Trustworthy AI, Non-Terrestrial Networks, Zero-Trust Security, Embedded Systems, and Advanced Digital Signal Processing.
\end{IEEEbiography}

\vfill

\end{document}